\documentstyle[eqsecnum,aps,twocolumn]{revtex}

\input{psfig}
\newcommand{\mybf}[1]{\mbox{\boldmath $#1$}}

\begin{document}
\draft

\def\be{\begin{equation}}
\def\ee{\end{equation}}
\def\beq{\begin{eqnarray}}
\def\eeq{\end{eqnarray}}

\title{
Effect of Exclusion of Double Occupancies in $t$-$J$ Model:
Extension of Gutzwiller Approximation
}
\author{Masao Ogata\thanks{Address from April 2000, Department of Physics,
University of Tokyo, Hongo, Bunkyo-ku 113-0033 Japan.  
email: ogata@phys.s.u-tokyo.ac.jp} 
and Akihiro Himeda$^*$}
\address{Department of Basic Science, The University of Tokyo\\
3-8-1 Komaba, Meguro-ku, Tokyo 153-8902, Japan
}
%\inst{Department of Basic Science, The University of Tokyo\\
%3-8-1 Komaba, Meguro-ku, Tokyo 153-8902, Japan
%}
\date{\today}
\maketitle

\def\xx{\chi}
\def\dd{\Delta}

\begin{abstract}
A new type of analytic estimation of the effect of strong correlation 
is developed for the two-dimensional $t$-$J$ model.   
It is based on the Gutzwiller approximation which gives the 
renormalization of parameters, $t$ and $J$, due to the Gutzwiller's 
projection operator excluding the double occupancies.  
The finite-range correlations are taken into account compared 
with the conventional Gutzwiller approximation where only the on-site 
expectation values are considered.  
It is shown that the essential point of the renormalization is 
its nonlinear dependence on the expectation values of Cooper pairs 
and antiferromagnetic moment.  
In particular the renormalization factor of $J$ 
becomes anisotropic in the presence of antiferromagnetic moment, 
i.e., that for the $z$-component is enhanced compared with that for
the $xy$-component.  
The physical origin of this enhancement is identified as the 
surrounding antiferromagnetic correlations of a bond.  
The self-consistency equations for the uniform variational wave functions 
are derived and solved numerically.  
Our result gives a reasonable estimate of antiferromagnetic order 
parameters near half filling,
in contrast to the conventional slave-boson mean-field theory and 
the original Gutzwiller approximation.  
It is also found that, at half filling, the renormalization of  
$J$ represents some 
of the quantum fluctuations of the Heisenberg spin system in a different 
manner from the spin-wave theory.  
For finite doping, our results have some similarity 
to SO(5) symmetric theory.  
Applications to the inhomogeneous systems such as the vortex state, 
around nonmagnetic impurities, and stripe state
are discussed.   
\end{abstract}

\narrowtext

%\begin{document}
%\sloppy
%\maketitle

% body of paper here
\section{Introduction}

For understanding the basic physics of high-$T_{\rm c}$ superconductivity, 
the two-dimensional $t$-$J$ model has been extensively studied 
as one of the most promising and simplified models 
which describe charge and spin 
dynamics in the CuO$_2$ plane.\cite{Anderson,ZR}  
Mean-field theories\cite{KL,Suzu,ZGRS,Tanamoto} and numerical 
calculations\cite{YSS,Gros,Dagotto,YO}
have shown that the d$_{x^2-y^2}$ wave superconductivity 
(SC) takes place in a reasonable parameter region in the 
phase diagram ($J/t=0.3$ and the doping rate $\delta < 0.3$).  
Therefore, as far as the d-wave SC is concerned, the experimentally 
observed phase diagram of high-$T_{\rm c}$ superconductors and the 
results in the $t$-$J$ model agree quite well.  

Recently, however, there are some experiments which indicate the importance 
of the interplay between the antiferromagnetism and d-wave SC.  
The most interesting phenomenon is a realization of stripe state in 
some materials, which has charge ordering as well as 
incommensurate antiferromagnetic (AF) spin 
ordering.\cite{Tranquada,Kimura99,Wakimoto99}  
Other potential problems are the AF state induced around 
nonmagnetic impurities and a possibility of AF vortex 
cores.\cite{SO5,Arovas,lasv,ISS99}

In order to study these problems, it is 
necessary to develop a theory in which the d-wave SC and the 
AF correlation are treated in a reliable way.  
Usual mean-field theories generally 
overestimate the AF long-range order, so that they give unphysical 
results even in the uniform case.  
In this paper we develop a new type of analytic theory which gives a 
reliable estimate of the AF correlation as well as the d-wave SC.  
Our scheme is an extension of the Gutzwiller approximation.\cite{ZGRS,Ogawa}

In the $t$-$J$ model the double occupancy of up- and down-spin electrons 
at any site is prohibited.  
To study its ground state, it is natural to use a projected mean-field 
wave function in which the double occupancies are excluded.\cite{Anderson}  
As the mean-field wave functions, a SDW mean-field solution 
at half filling,\cite{YS2} BCS wave functions,\cite{YSS,Gros}
and a coexistent state of the AF and d-wave SC\cite{Chen,GL,Himeda} 
have been used.  
Variational energies are calculated in the variational Monte Carlo 
(VMC) simulations, 
which treat exactly the constraints of no double occupancies.  
Among the above trial states, 
the coexistent state has the lowerest variational energy in the 
doping region $\delta< 0.1$.\cite{Chen,GL,Himeda}

On the other hand, the analytic theories in which 
the constraint is treated approximately give large discrepancies in 
the estimation of the AF correlation.  
For example, in the slave-boson mean-field theory 
which takes account of the constraint as an average, 
the AF order is too overestimated 
and it extends up to unphysical doping rates ($\delta<0.2$).\cite{Inaba}  
Thus in the slave-boson theory, we are unable to discuss the 
stripe state, for example, 
because it is stabilized near $\delta=0.125$ doping where the slave-boson 
theory gives the AF state even for the bulk.  

The Gutzwiller approximation is more advanced approximation 
than the simple mean-field theories in treating the constraint.
Renormalizations of expectation values are introduced 
by comparing the statistical weighting factors in the wave functions 
with and without projection.\cite{ZGRS,Vollhardt}   
As a result, the parameters $t$ and $J$ are renormalized to 
$g_t t$ and $g_s J$.  
It has been shown that the Gutzwiller approximation gives a fairly 
reliable estimation 
of the variational energy for the pure d-wave SC state when it is 
compared with the VMC results.\cite{ZGRS,YO}  
However it was shown that there is no region in the phase diagram where the 
AF state is stabilized.\cite{ZGRS} 
This contradicts with the VMC results.  

In order to clarify the origin of these discrepancies, 
we investigated the VMC data\cite{Himeda}
and found that the Gutzwiller approximation has to 
be modified when the AF correlations are present.  
Based on these observations, we extend the Gutzwiller approximation 
in this paper and derive an analytic 
formalism for the renormalization factors which reproduces the 
VMC results.  
We show that 
it is important to take account of the longer-range correlations 
for the weighting factors, 
in contrast to the previous approximation where only the site-diagonal 
expectation values such as $\langle n_{i\sigma} \rangle$ are considered.  
As a result, the renormalization factors for $t$ and $J$ have 
nonlinear dependences on the expectation values of Cooper pairs 
and AF moment.  
We think that this is the essence of the 
strong correlation in the sense that the renormalization 
appears solely from the projection operators.  

The merit of the present scheme is that it 
can be easily applied to the inhomogeneous systems, where 
the VMC simulations are not so feasible.  
Thus a reliable analytic formulation can be given to the problems, such 
as the stripe state, vortex cores and states around impurities, where 
the interplay between the AF and d-wave SC plays an important role.  
Preliminary results of these problems have been described 
elsewhere.\cite{lasv,ISS99,Yasuoka}  

The outline of this paper is as follows: In Sec.\ II we briefly review the 
formulation of the original Gutzwiller approximation in order to 
make the present paper self-contained.  
We also show our final results for the extended Gutzwiller 
approximation before going into the details of the derivation.  
In Sec.\ III, the original Gutzwiller approximation is extended 
in order to include longer-range correlations.  
We divide the whole system into cells and introduce 
the configurations in each cell for evaluating the 
weighting factors.  
Using the general formulation developed in Sec.\ III, 
we obtain the renormalization factors for the half-filled 
case in Sec.\ IV.  
We approximate the weighting factors in a perturbative way with 
respect to the nearest-neighbor correlations.  
In IV.D, the physical origin of the enhancement of the 
renormalization factor for the $z$-component of exchange interaction 
is discussed in a viewpoint of statistical weights of real-space 
spin configurations.  
In Sec.\ V, the results at half filling are generalized to the 
less-than-half-filled case.  
The present method is applied to the projected 
variational states in Sec.\ VI.  
The self-consistency equations are derived and solved numerically 
to show that they give reasonable estimate of AF long-range order 
near half filling.  
Finally section VII is devoted to a summary and discussions on 
related problems.

\section{$t$-$J$ model and the original Gutzwiller approximation}

We consider the Gutzwiller approximation for the two-dimensional 
$t$-$J$ model on a square lattice:
\be
{\widehat{\cal H}}=-t\sum_{\langle ij\rangle\sigma}P_G(c_{i\sigma}^{\dagger}
 c_{j\sigma}+h.c.)P_G+J\sum_{\langle ij\rangle}
 \,{\mybf{S}}_{i}\cdot {\mybf{S}}_j,
\ee
where $\langle ij\rangle$ represents the sum over the nearest-neighbor sites
and ${\mybf{S}}_i=c_{i\alpha}^{\dagger}(\frac{1}{2}
{\mybf{\sigma}})_{\alpha\beta}c_{i\beta}$.
%{\sigma})_{\alpha\beta}c_{i\beta}$.
The Gutzwiller's projection operator $P_G$ is defined as
$P_G=\prod_j (1-\hat{n}_{j\uparrow}\hat{n}_{j\downarrow})$.

For this Hamiltonian, we assume the projected BCS-SDW mean-field 
wave function\cite{Himeda}  
\be
\label{vwf}
 |\psi\rangle=P_G|\psi_0(\Delta_d^{\rm V},\Delta_{\rm af}^{\rm V},
\mu^{\rm V})\rangle,
\ee
where $\Delta_d^{\rm V}$, $\Delta_{\rm af}^{\rm V}$ and $\mu^{\rm V}$ 
are the variational parameters 
relating to d-wave SC, AF and chemical potential, respectively, and 
$|\psi_0(\Delta_d^{\rm V}, \Delta_{\rm af}^{\rm V}, \mu^{\rm V})\rangle$ 
is a Hartree-Fock type wave function with 
d-wave SC and AF orders.
The wave function $|\psi\rangle$ is a natural generalization of 
the RVB state proposed by Anderson.\cite{Anderson}  
It has been shown that the coexistent state of AF and d-wave SC 
has the best variational energy in the VMC simulations for the 
doping rate $\delta<0.1$.\cite{Chen,GL,Himeda}  

In evaluating the variational energy, the projection operator makes 
difficulties for an analytic approach.  
The Gutzwiller approximation was developed for this purpose.  
In this method, the effect of the projection is taken into account 
by renormalizations of expectation values as follows:
\be
 \langle c_{i\sigma}^{\dagger} c_{j\sigma}\rangle =
 g_t\langle c_{i\sigma}^{\dagger} c_{j\sigma}\rangle_{0}, \quad
 \langle {\mybf{S}}_i\cdot{\mybf{S}}_j\rangle =
 g_s\langle {\mybf{S}}_i\cdot {\mybf{S}}_j\rangle_{0},  
\label{defgs}
\ee
where $\langle \cdots \rangle_0$ represents the expectation value 
in terms of 
$|\psi_0\rangle = 
|\psi_0(\Delta_d^{\rm V}, \Delta_{\rm af}^{\rm V}, \mu^{\rm V})\rangle$,
and $\langle \cdots \rangle$ represents the normalized expectation 
value in $|\psi \rangle = P_G|\psi\rangle_0$; 
\beq
\langle \widehat{\cal O} \rangle \equiv 
\frac{\langle \psi | \widehat{\cal O} | \psi\rangle }
{\langle \psi| \psi \rangle } 
=\frac{\langle \psi_0 | P_G \widehat{\cal O} P_G  | \psi_0\rangle }
{\langle \psi_0 | P_G P_G  | \psi_0\rangle } .
\eeq
The coefficients $g_t$ and $g_s$ are the renormalization factors 
of the expectation values, which we call as Gutzwiller factors 
in the following.  
Using these notations, the variational energy 
$E_{\rm var} = \langle {\widehat{\cal H}}\rangle$
is rewritten as 
\be
E_{\rm var} = \langle {\widehat{\cal H}}_{\rm eff} \rangle_0,
\label{Evar}
\ee
where the parameters $t$ and $J$ in $\widehat{\cal H}$ are 
replaced with 
\be
t_{\rm eff} = g_t t, \qquad J_{\rm eff}=g_s J.  
\ee

A systematic estimation of $g_t$ and $g_s$ 
was developed by Ogawa {\it et al.}\cite{Ogawa} whose method is 
briefly reviewed in II.A.  For the case with AF order parameter, 
it can be shown that 
\be
      g_t=\frac{2\delta(1-\delta)}{1-\delta^2+4m^2}, 
\quad g_s=\frac{4(1-\delta)^2}{(1-\delta^2+4m^2)^2},
\label{origG}
\ee
where $\delta$ is the hole concentration, $\delta=1-n$, and $m$ is 
the expectation value of AF order parameter in terms of 
$|\psi_0\rangle$ defined as
\be
m=\frac{(-1)^j}{2}\bigl( 
\langle {\hat n}_{j\uparrow}\rangle_0 - \langle {\hat n}_{j\downarrow}
\rangle_0 \bigr).
\ee
These Gutzwiller factors, however, do not reproduce the VMC results, 
as mentioned in I.  We find that this discrepancy is due to the 
fact that only the site-diagonal expectation values 
are taken into account in obtaining Eq.\ (\ref{origG}).  
In this paper we extend the method by 
Ogawa {\it et al.} systematically to include the longer range correlations. 
We show that it is important to include the effects of the 
nearest-neighbor expectation values, such as 
\beq
\xx &&=\langle c_{i\sigma}^\dagger c_{j\sigma} \rangle_0, \nonumber\\
\dd &&=\langle c_{i\uparrow}^\dagger c_{j\downarrow}^\dagger \rangle_0,
\label{defxx}
\eeq
in the estimation of $g_t$ and $g_s$.  
This corresponds to taking account of the effect of 
surroundings of the corresponding bond.  

Furthermore we find that the Gutzwiller factors $g_s$ 
for the exchange interaction have different values for the $z$ component 
(denoted as $g_s^Z$) and $xy$ component ($g_x^{XY}$); i.e.,
\beq
 \langle S^z_i S^z_j \rangle &&= g_s^Z \langle S^z_i S^z_j \rangle_{0},  
\nonumber\\
 \langle S^+_i S^-_j \rangle &&= g_s^{XY} \langle S^+_i S^-_j \rangle_{0},
\eeq
instead of (\ref{defgs}) in the presence of $m$.  
Actually from the VMC calculations,\cite{Himeda} 
we have estimated the behaviors of $g_s^Z$ and $g_s^{XY}$ 
as a function of $m$, using the relations
\be
\label{VMCGF}
g_s^Z= \frac{\langle S^z_i S^z_j \rangle}{\langle S^z_i S^z_j \rangle_{0}},
\qquad 
g_s^{XY}= \frac{\langle S^+_i S^-_j \rangle}{\langle S^+_i S^-_j \rangle_{0}},
\ee
where the numerators $\langle S^z_i S^z_j \rangle$ and 
$\langle S^+_i S^-_j \rangle$ are evaluated numerically in VMC 
simulations.  
It was found that $g_s^Z$ is enhanced compared with $g_s^{XY}$ 
and that this enhancement is essential for the 
stabilization of the AF order.  
In the present extended Gutzwiller approximation, this result is reproduced. 
Physically the weighting factor of $g_s^Z$ 
for a specific configuration of a bond 
is enhanced due to the effect of the surrounding AF correlations.  

Before going into details of the derivation, we show our final results.  
They are summarized as follows:
\beq
g_s^{XY} = \biggl( \frac{2(1-\delta)}{1-\delta^2+4m^2}\biggr)^2 a^{-7},
\eeq
and
\be
g_s^Z = g_s^{XY} \frac{1}{4m^2+X_2} \biggl[
X_2 + 4m^2 \bigl\{ 1+ \frac{6 X_2 (1-\delta)^2}{1-\delta^2+4m^2} 
a^{-3} \bigr\}^2 \biggr], 
\label{gszfinal}
\ee
where 
\beq
a &&=1+\frac{4X}{(1-\delta^2+4m^2)^2}, \nonumber\\
X &&=2\delta^2(\dd^2-\xx^2)+8m^2(\xx^2+\dd^2)+4(\xx^2+\dd^2)^2, \nonumber\\
X_2 &&=2(\xx^2+\dd^2).
\label{XX2}
\eeq
The above expressions sufficiently reproduce the results obtained 
in the VMC simulations.  
However we find that the slight difference can be improved by using 
\be
X = 2\delta^2(\dd^2-\xx^2)+8m^2(\xx^2+\dd^2)+2(\xx^2+\dd^2)^2, 
\ee
instead of Eq.\ (\ref{XX2}).  
A typical $m$-dependence of $g_s^Z$ and $g_s^{XY}$ are shown in Fig.\ 1
for the half-filled case, $\delta=0$, and with $\Delta$ being fixed 
at the two values, $\Delta=0.02$ and $0.18$.  
The behaviors in Fig.\ 1 agree with those obtained in the VMC 
simulations using the relation in Eq.\ (\ref{VMCGF}).  
The meanings of the rather complicated Gutzwiller factors become 
apparent in the following sections.  

\begin{figure}
\psfig{figure=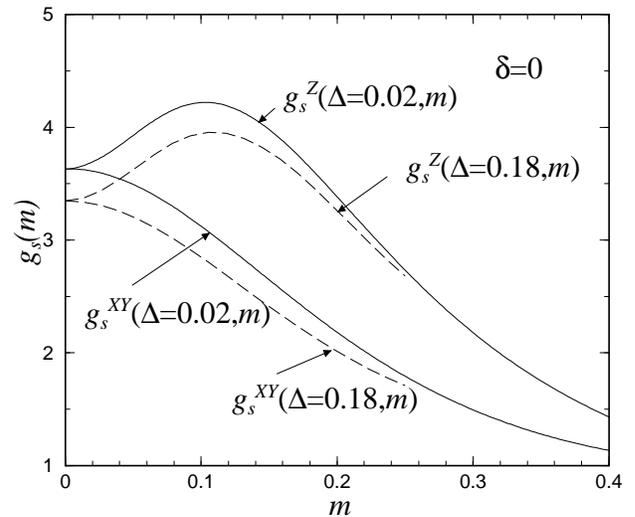,height=7cm}
\caption{Typical $m$-dependences of the Gutzwiller's renormalization 
factors, $g_s^Z$ and $g_s^{XY}$ for the half-filled case.  
$g_s^Z$ ($g_s^{XY}$) is the Gutzwiller factor for the 
$z$- ($xy$-) component of the exchange interaction, respectively.  
$\Delta$ is the nearest-neighbor expectation value of 
$\langle c_{i\uparrow}^\dagger c_{j\downarrow}^\dagger \rangle_0$ 
in the wave function without the projection.  It is fixed
at values $\Delta=0.02$ and $0.18$.  
$m$ is the expectation value of $1/2(n_{i\uparrow}-n_{i\downarrow})$ 
without the projection.  
\label{fig:1}}
\end{figure}

The Gutzwiller factor for the hopping term is given by
\beq
g_t=\frac{2\delta (1-\delta)}{1-\delta^2+4m^2} \ 
\frac{(1+\delta)^2-4m^2-2X_2}{(1+\delta)^2-4m^2}a.
\label{gtfinal}
\eeq
Figure 2 shows the $m$-dependences of $g_s^{XY}, g_s^Z$ and 
$g_t$ for $\delta=0.12$, which can be compared with the 
VMC results in Ref.\ \cite{Himeda}.  
Note here that the slave-boson mean-field theory simply gives 
$g_t = \langle b_i^\dagger b_j \rangle = \delta$ and $g_s=1$.  
The nonlinear dependences of $g_s^{XY}, g_s^Z$ and $g_t$ on 
$\dd, \xx$ and $m$ are beyond the slave boson theory.  
In the following subsection we review the original formulation 
of the Gutzwiller approximation which is useful for the generalization 
in the later sections.  
\begin{figure}
\psfig{figure=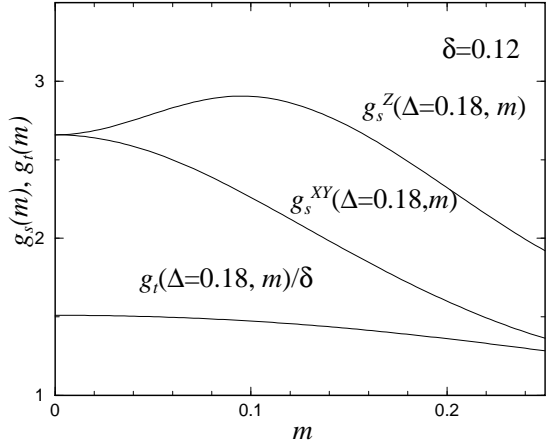,height=6cm}
\caption{Typical $m$-dependences of $g_s^Z, g_s^{XY}$ and $g_t/\delta$
for the doping rate $\delta=0.12$ in the $t$-$J$ model.  
The meanings of the parameters are the same as in Fig.\ 1.
\label{fig:2}}
\end{figure}

\subsection{Original formulation of Gutzwiller approximation}

Let us start with the formulation of the Gutzwiller 
approximation developed by Ogawa {\it et al} \cite{Ogawa}
when it is applied to the $t$-$J$ model.  
In this subsection we do not consider the AF order.  
The ultimate aim is to evaluate the variational energy 
\be
E_{\rm var} = \langle {\widehat{\cal H}}\rangle 
=\frac{\langle \psi |\widehat{\cal H}|\psi\rangle}
{\langle \psi | \psi\rangle}, 
\ee
using $|\psi\rangle = P_G|\psi_0\rangle$.  
First, Ogawa {\it et al} rewrote the denominator as ($P_G^2=P_G$)
\beq
\langle \psi | \psi\rangle 
&&=\langle \psi_0 | P_G P_G | \psi_0\rangle \nonumber\\
&&=\langle \psi_0 | \prod_j (1-\widehat n_{j\uparrow} \widehat n_{j\downarrow})
| \psi_0\rangle  \nonumber\\
&&=\langle \psi_0 | \prod_j \{
 \widehat n_{j\uparrow}(1-\widehat n_{j\downarrow}) 
+\widehat n_{j\downarrow}(1-\widehat n_{j\uparrow}) \nonumber\\
&&\qquad\qquad +(1-\widehat n_{j\uparrow})(1-\widehat n_{j\downarrow})
\} | \psi_0\rangle .
\label{deno}
\eeq
The expansion of the product in (\ref{deno})  
leads to various real-space configurations.  
Therefore $\langle \psi | \psi\rangle$
can be rewritten as
\beq
\sum_{\rm config.}
\langle \psi_0 &&| 
\prod_{j\in {\cal A}} \widehat n_{j\uparrow}(1-\widehat n_{j\downarrow}) 
\prod_{j'\in {\cal B}} \widehat n_{j'\downarrow}(1-\widehat n_{j'\uparrow}) 
\nonumber\\
&&\times 
\prod_{j''\in {\cal E}} (1-\widehat n_{j''\uparrow})
(1-\widehat n_{j''\downarrow})
| \psi_0\rangle ,
\label{deno2}
\eeq
where $\cal A$ ($\cal B$) is a subset of the lattice sites which is singly 
occupied by an up-spin (down-spin) electron and $\cal E$ is a subset 
which is vacant.  
The summation is over all the possible configurations, i.e., all the 
possible combinations of $\{{\cal A}, {\cal B}, {\cal E}\}$.

In principle, Eq.\ (\ref{deno2}) can be calculated using Wick's theorem 
since $|\psi_0\rangle$ is a mean-field wave function, although 
it is very complicated.  
The simplest Gutzwiller approximation is to estimate (\ref{deno2}) 
by using only the site-diagonal expectation values, which gives 
\beq
\sum_{\rm config.}
&&\prod_{j\in {\cal A}} 
\langle \widehat n_{j\uparrow}(1-\widehat n_{j\downarrow}) \rangle_0
\prod_{j'\in {\cal B}} 
\langle \widehat n_{j'\downarrow}(1-\widehat n_{j'\uparrow}) \rangle_0
\nonumber\\
\times
&&\prod_{j''\in {\cal E}} 
\langle (1-\widehat n_{j''\uparrow})(1-\widehat n_{j''\downarrow}) \rangle_0,
\label{deno3}
\eeq
with $\langle \cdots \rangle_0$ meaning the expectation value 
in $|\psi_0\rangle$.  

For example, the contribution from the site $j\in {\cal A}$ is 
evaluated as 
\be
\omega_A \equiv 
\langle \widehat n_{j\uparrow}(1-\widehat n_{j\downarrow}) \rangle_0 =
\frac{n}{2}\bigl( 1-\frac{n}{2} \bigr),
\ee
where $n$ is the average electron density, $n=N_e/N$, with $N$ ($N_e$) 
being the total number of sites (total electron number).
We call $\omega_A$ as the weight for $j$ site belonging to the 
subset ${\cal A}$.  
Since the hole density is $\delta = 1-n$, we obtain
\beq
\omega_A &&= \omega_B = \frac{n}{2}\left( 1-\frac{n}{2} \right)  
= \frac{1-\delta^2}{4}, \nonumber\\
\omega_E &&= \left( 1-\frac{n}{2} \right)^2 
= \frac{(1+\delta)^2}{4}. 
\eeq

Then the expectation value (\ref{deno3}) is rewritten as 
\be 
\label{deno0}
\sum_{\rm config.} \omega_A^{N_A} \omega_B^{N_B} \omega_E^{N_E} 
=\frac{N!}{N_A! N_B! N_E!}\omega_A^{N_A} \omega_B^{N_B} \omega_E^{N_E},
\ee
where $N_A$ is the number of sites in the subset $\cal A$, and so on,
($N_A+N_B+N_E=N$).
In this simple case,
$N_A$ is equal to the number of up-spin electrons, so that 
\be
N_A=N_B=\frac{N_e}{2}, \qquad N_E=N-N_e.
\ee

A similar procedure can be carried out for the estimation of 
the numerator in $E_{\rm var}$.  For the exchange term, we have 
\beq
\langle \psi_0&&| P_G S_\ell^+ S_m^- P_G | \psi_0\rangle \nonumber\\
&&=\langle \psi_0 | S_\ell^+ S_m^- 
\prod_{j\ne \ell, m} \{
 \widehat n_{j\uparrow}(1-\widehat n_{j\downarrow}) 
+\widehat n_{j\downarrow}(1-\widehat n_{j\uparrow}) \nonumber\\
&&\qquad\qquad\quad \quad
+(1-\widehat n_{j\uparrow})(1-\widehat n_{j\downarrow}) \} 
| \psi_0\rangle \nonumber\\
&&=\sum_{\rm config.}
\langle \psi_0 | S_\ell^+ S_m^- 
\prod_{j \in {\cal A'}} \widehat n_{j\uparrow}(1-\widehat n_{j\downarrow}) 
\prod_{j'\in {\cal B'}} \widehat n_{j'\downarrow}(1-\widehat n_{j'\uparrow}) 
\nonumber\\
&&\qquad\qquad\qquad\times 
\prod_{j''\in {\cal E'}} (1-\widehat n_{j''\uparrow})
(1-\widehat n_{j''\downarrow})
| \psi_0\rangle .
\label{nume}
\eeq
%\beq
%\langle \psi_0| P_G S_\ell^+ S_m^- P_G | \psi_0\rangle
%&&=\langle \psi_0 | S_\ell^+ S_m^- 
%\prod_{i\ne \ell, m} \{
% \widehat n_{i\uparrow}(1-\widehat n_{i\downarrow}) 
%+\widehat n_{i\downarrow}(1-\widehat n_{i\uparrow}) 
%+(1-\widehat n_{i\uparrow})(1-\widehat n_{i\downarrow}) \} 
%| \psi_0\rangle \nonumber\\
%&&=\sum_{\rm config.}
%\langle \psi_0 | S_\ell^+ S_m^- 
%\prod_{i\in {\cal A'}} \widehat n_{i\uparrow}(1-\widehat n_{i\downarrow}) 
%\prod_{i\in {\cal B'}} \widehat n_{i\downarrow}(1-\widehat n_{i\uparrow}) 
%\prod_{i\in {\cal E'}} (1-\widehat n_{i\uparrow})(1-\widehat n_{i\downarrow})
%| \psi_0\rangle ,
%\label{nume}
%\eeq
Here the summation is over all the possible configurations 
$\{{\cal A}', {\cal B}', {\cal E}'\}$ in which the two sites 
$\ell$ and $m$ are excluded. 
The evaluation of (\ref{nume}) with the site-diagonal expectation values for 
$j\ne \ell, m$ leads to 
\be
\frac{(N-2)!}{N_{A'}! N_{B'}! N_{E'}!}\omega_A^{N_{A'}} \omega_B^{N_{B'}} 
\omega_E^{N_{E'}} \langle S_\ell^+ S_m^- \rangle_0 .
\label{GAexp}
\ee
Since the sites $\ell$ and $m$ contain one up-spin electron and 
one down-spin electron, we have
$N_{A'}=\frac{N_e}{2} -1$, $N_{B'}=\frac{N_e}{2} -1$, and 
$N_{E'}=N_E=N-N_e$.  

By combining the estimation of the denominator in Eq.\ (\ref{deno0}), 
we obtain the original 
Gutzwiller approximation 
\beq
\langle S_\ell^+ S_m^- \rangle &&= 
\frac{\langle \psi_0 | P_G S_\ell^+ S_m^- P_G | \psi_0\rangle} 
{\langle \psi_0 | P_G P_G | \psi_0\rangle} \nonumber \\
&&=\frac{\frac{N_e}{2}\frac{N_e}{2}}{N(N-1)}\omega_A^{-1}\omega_B^{-1}
\langle S_\ell^+ S_m^- \rangle_0 \nonumber \\
&&=\frac{4}{(1+\delta)^2}
\langle S_\ell^+ S_m^- \rangle_0 .
\eeq
This gives $g_s^{XY}=4/(1+\delta)^2$.  

\subsection{Antiferromagnetic case}

Ogawa {\it et al}\cite{Ogawa} also considered the above formalism in 
the case with AF long-range order for the Hubbard model.  
The application to the $t$-$J$ model was carried out by 
Zhang {\it et al}.\cite{ZGRS}   
In this subsection we follow their methods to obtain the Gutzwiller 
approximation in the $t$-$J$ model. 

In the AF case, the sublattices 1 and 2 are 
distinguished and their magnetizations are defined as
\be
\frac{1}{2} \left( \langle \widehat n_{j\uparrow} \rangle_0 -
\langle \widehat n_{j\downarrow} \rangle_0 \right) =m,
\ee
for the sublattice 1 and $-m$ 
%\be
%\frac{1}{2} \left( \langle \widehat n_{i\uparrow} \rangle_0 -
%\langle \widehat n_{i\downarrow} \rangle_0 \right) =-m,
%\ee
for the sublattice 2.  Thus we denote 
\beq
\langle \widehat n_{j\uparrow} \rangle_0 &&= \frac{n}{2}+m \equiv r, 
\nonumber\\
\langle \widehat n_{j\downarrow} \rangle_0 &&= \frac{n}{2}-m \equiv w,
\eeq
for the sublattice 1, where $r$ ($w$) means the average electron 
density with the right (wrong) spin direction on the sublattice.  
$r$ and $w$ are exchanged for the sublattice 2.

Using these notations, the denominator of $E_{\rm var}$ is 
evaluated as
\beq
\langle \psi_0 | P_G P_G | \psi_0\rangle 
=&&\sum_{\rm config.}
\omega_{A1}^{N_{A1}} \omega_{B1}^{N_{B1}} \omega_{E1}^{N_{E1}}
\omega_{A2}^{N_{A2}} \omega_{B2}^{N_{B2}} \omega_{E2}^{N_{E2}} \nonumber\\
=&&\sum_{N_{A1}, N_{B1}}
\frac{\left(\frac{N}{2}\right) !}{N_{A1}! N_{B1}! N_{E1}!}
\frac{\left(\frac{N}{2}\right) !}{N_{A2}! N_{B2}! N_{E2}!} \nonumber\\
&&\times
\omega_{A1}^{N_{A1}} \omega_{B1}^{N_{B1}} \omega_{E1}^{N_{E1}}
\omega_{A2}^{N_{A2}} \omega_{B2}^{N_{B2}} \omega_{E2}^{N_{E2}},
\label{denoaf}
\eeq
where the configurations in the sublattice 1 are specified by 
$(N_{A1}, N_{B1}, N_{E1})$, and so on.  
Contrary to the previous subsection, Eq.\ (\ref{denoaf}) has a 
summation over possible values of $N_{A1}$ and $N_{B1}$ because 
there are only four constraints between six numbers, 
$(N_{A1}, N_{B1}, N_{E1}, N_{A2}, N_{B2}, N_{E2})$:
\beq
N_{A1}+N_{B1}+N_{E1}&&=\frac{N}{2}, \quad N_{A2}+N_{B2}+N_{E2}=\frac{N}{2}, 
\nonumber \\
N_{A1}+N_{A2}&&=\frac{N_e}{2},  \qquad N_{B1}+N_{B2}=\frac{N_e}{2}.
\nonumber \\
\eeq
The weights in the site-diagonal expectation values are
\beq
\omega_{A1} &&= \langle \widehat n_{j\uparrow} \rangle_0 
\langle 1-\widehat n_{j\downarrow} \rangle_0 = r(1-w), \nonumber \\
\omega_{B1} &&= \langle \widehat n_{j\downarrow} \rangle_0 
\langle 1-\widehat n_{j\uparrow} \rangle_0 = w(1-r), \nonumber \\
\omega_{E1} &&= \langle 1-\widehat n_{j\uparrow} \rangle_0 
\langle 1-\widehat n_{j\downarrow} \rangle_0 = (1-r)(1-w),
\eeq
for the sublattice 1, and 
\beq
\omega_{A2}&&=\omega_{B1}, \nonumber \\
\omega_{B2}&&=\omega_{A1}, \nonumber \\
\omega_{E2}&&=\omega_{E1},
\eeq
for the sublattice 2.

Since $N_{A1}, N_{B1}$ and $N$ are large numbers, Eq.\ (\ref{denoaf}) is 
further approximated by the largest term in the summation.  
By taking the partial derivative of the logarithm of each term in 
(\ref{denoaf}) with respect to $N_{A1}$, we obtain 
\be
-\ln \overline{N_{A1}} + \ln (\frac{N_e}{2}-\overline{N_{A1}}) 
+ \ln \omega_{A1} - \ln \omega_{A2}=0, 
\label{maximize}
\ee
with $\overline{N_{A1}}$ being the value of $N_{A1}$ which gives 
the largest term in the summation.  By solving (\ref{maximize}), 
$\overline{N_{A1}}$ is given by 
\be
\overline{N_{A1}}=\frac{r(1-w)}{r(1-w)+w(1-r)} \frac{N_e}{2} 
=\frac{r(1-w)}{n-2rw} \frac{N_e}{2} .
\label{maximizesol}
\ee
Similarly we obtain 
\beq
\overline{N_{B1}} = \frac{w(1-r)}{n-2rw} \frac{N_e}{2} .
\eeq
The summation in (\ref{denoaf}) is approximated by the term with 
$\overline{N_{A1}}$ and $\overline{N_{B1}}$.  

Again the numerator such as 
$\langle \psi_0 | P_G S_\ell^+ S_m^- P_G | \psi_0\rangle$ is 
approximated in the similar way.  As a result, the Gutzwiller 
approximation in the presence of AF order, $m$,
becomes
\beq
\langle S_\ell^+ S_m^- \rangle&&= 
\frac{\langle \psi_0 | P_G S_\ell^+ S_m^- P_G |\psi_0\rangle} 
{\langle \psi_0 | P_G P_G | \psi_0\rangle} \nonumber\\
&&=\frac{\overline{N_{A1}}\  \overline{N_{B1}}}{\frac{N}{2}\frac{N}{2}} 
\omega_{A1}^{-1}\omega_{B1}^{-1}
\langle S_\ell^+ S_m^- \rangle_0 \nonumber \\
&&=\frac{n^2}{(n-2rw)^2} \langle S_\ell^+ S_m^- \rangle_0 \nonumber \\
&&=\frac{4(1-\delta)^2}{(1-\delta^2+4m^2)^2} 
\langle S_\ell^+ S_m^- \rangle_0 ,
\label{gs0}
\eeq
where a useful relation 
\be
\frac{n}{n-2rw}=\frac{2(1-\delta)}{1-\delta^2+4m^2}
\ee
has been used.  Equation (\ref{gs0}) gives 
\be
g_s^{XY}=\frac{4(1-\delta)^2}{(1-\delta^2+4m^2)^2},
\ee
which is simply Eq.\ (\ref{origG}).  

It is straightforward to see
\be
g_s^{Z} = g_s^{XY},
\ee
and 
\be
g_t = \frac{2\delta(1-\delta)}{1-\delta^2+4m^2}.
\ee

\section{Extension of the Gutzwiller approximation: Formulation}
%\subsection{Formalism}

The approximation in the preceding section is restricted 
to the site-diagonal (on-site) expectation values.  
It was shown that this approximation 
is not good enough to reproduce the AF order obtained 
in the VMC simulation.\cite{ZGRS} 
A straightforward extension of the previous Gutzwiller approximation is to 
consider longer-range expectation values.  
The nearest-neighbor expectation was considered before.\cite{Hsu,Sigrist}
In this section, 
we develop a general formalism to take account of the longer-range 
effects systematically.  

In evaluating the denominator of $E_{\rm var}$, i.e.,
$\langle \psi_0 | P_G P_G  | \psi_0\rangle$, various configurations 
have been classified depending on the state ($\uparrow, \downarrow$ or a hole)
on each site as $\{{\cal A}, {\cal B}, {\cal E}\}$
in the previous section.  
Instead of this we divide the whole system into {\it cells} 
consisting of $N_c$ sites.  (Later we take $N_c$ as a large enough number.)
For each cell, there are $3^{N_c}$ configurations because each site 
has $\uparrow, \downarrow$ or a hole.  
We denote these configurations as {\it states} of a cell. 
Using these states, 
the denominator of $E_{\rm var}$ can be approximated as
\beq
\langle \psi_0 | P_G P_G  | \psi_0\rangle 
&&= \sum_{\rm all\ the\ possible\ states} 
\prod_{i=1}^K \omega_i^{N_i} \nonumber\\
&&= \sum_{\{ N_i\} } \frac{\bigl(\frac{N}{N_c}\bigr)!}{\prod_{i=1}^K(N_i)!}
\prod_{i=1}^K \omega_i^{N_i}, 
\label{denogen}
\eeq
with $K = 3^{N_c}$.
Here $N_i$ is the number of cells in the $i$-th state, and 
$\omega_i$ is the weight of the $i$-th state cell 
in the analogy with Eq.\ (\ref{denoaf}).  
Explicitly the weight $\omega_i$ is defined as 
\beq
\omega_i =&& \langle \psi_0 | 
\prod_j    \widehat n_{j\uparrow}(1-\widehat n_{j\downarrow}) 
\prod_{j'} \widehat n_{j'\downarrow}(1-\widehat n_{j'\uparrow}) 
\nonumber\\
&&\times 
\prod_{j''} (1-\widehat n_{j''\uparrow})(1-\widehat n_{j''\downarrow})
| \psi_0\rangle ,
\eeq
with $j, j'$ and $j''$ being the sites in the cell.  
To calculate $\omega_i$ is the most important part of the present theory
which we carry out later.  

The last summation in (\ref{denogen}) 
is over all the possible values of $N_i$ under the following constraints:
\def\nhi{{n_{{\rm h} i}}}
\def\nhiz{{n_{{\rm h} i_0}}}
\beq
&&\sum_{i=1}^K n_{\uparrow i} N_i = \frac{N_e}{2}, \qquad
\sum_{i=1}^K n_{\downarrow i} N_i = \frac{N_e}{2}, \nonumber\\
&&\sum_{i=1}^K \nhi N_i = N- N_e, 
\label{constr}
\eeq
where $n_{\uparrow i}$ ($n_{\downarrow i}$) represents 
the number of up-spins (down-spins) in the $i$-th state cell, and
$\nhi$ represents the number of holes, respectively.  
Since a cell has $N_c$ sites, a relation 
\beq 
n_{\uparrow i} + n_{\downarrow i} + \nhi = N_c,
\label{constr2}
\eeq
holds.  

Under the above constraints, we look for the largest term 
in Eq.\ (\ref{denogen}) 
as was done in the original Gutzwiller approximation 
(see Eq.\ (\ref{maximize})).  
Since the constraints (\ref{constr}) are rather complicated, it is 
useful to introduce 
the Lagrange multipliers ($\mu_\uparrow, \mu_\downarrow, \lambda$).  
Taking the partial derivative of 
\beq
&&\ln \biggl[ \frac{\bigl(\frac{N}{N_c}\bigr)!}{\prod_{i=1}^K(N_i)!}
\prod_{i=1}^K \omega_i^{N_i} \biggr] 
-\mu_\uparrow 
\bigl(\sum_{i=1}^K n_{\uparrow i} N_i - \frac{N_e}{2} \bigr) \nonumber\\
&&-\mu_\downarrow 
\bigl(\sum_{i=1}^K n_{\downarrow i} N_i -\frac{N_e}{2} \bigr) 
-\lambda \bigl( \sum_{i=1}^K \nhi N_i - N + N_e \bigr),
\eeq
we obtain 
\beq
\overline{N_i} = \omega_i {\rm exp} (-\mu_\uparrow n_{\uparrow i}
-\mu_\downarrow n_{\downarrow i} - \lambda \nhi).
\eeq
From symmetry we find $\mu_\uparrow =\mu_\downarrow =\mu$, and using 
the relation (\ref{constr2}) we get
\beq
\overline{N_i} = \omega_i {\rm e}^{-\mu N_c} {\rm e}^{(\mu-\lambda)\nhi}.
\eeq
Furthermore we introduce new variables $W$ and $p$ as
\beq
{\rm e}^{-\mu N_c} &&\equiv \frac{N}{N_c} \frac{1}{W}, \nonumber\\
{\rm e}^{\mu-\lambda} &&\equiv p.
\eeq
Using these variables, $\overline{N_i}$ is rewritten as
\beq
\overline{N_i} = \frac{N}{N_c} \frac{\omega_i}{W} p^\nhi.
\eeq

\def\MYdelta{{\delta}}

The variables $W$ and $p$ are to be determined from the constraints
(\ref{constr}) which are equivalent to 
\beq
\sum_{i=1}^K \overline{N_i} = \frac{N}{N_c},  \quad
\sum_{i=1}^K \nhi \overline{N_i} =N- N_e.
\eeq
These conditions become
\beq
\sum_{i=1}^K \frac{\omega_i}{W} p^\nhi &&= 1,  \nonumber\\
\sum_{i=1}^K \nhi \frac{\omega_i}{W} p^\nhi &&=\MYdelta N_c.
\eeq
%where we have introduced $\MYdelta=1-N_e/N$ in order to distinguish it 
%from the expectation value $\delta=1-\langle n_i \rangle_0$.  
It is convenient to classify all the possible states of a cell  
into subgroups which contain $j$ holes ($j= 0, 1, 2, \cdots, N_c$).  
(We call these subgroups as $j$-hole sectors.)  
Then the quantity 
\beq
W_j = \sum_{i\ {\rm with}\ j \ {\rm holes}} \omega_i, 
\label{wj}
\eeq
represents the total weight of the states in this subgroup.  
Using $W_j$, we can rewrite the constraints as 
\beq
\sum_{j=0}^{N_c} \frac{W_j}{W} p^j &&= 1, \nonumber\\
\sum_{j=0}^{N_c} j\frac{W_j}{W} p^j &&= \MYdelta N_c.
\label{constr3}
\eeq
The first equality is rewritten as $W=\sum_j W_j p^j$ so that 
$W$ represents the total weight.  
In the following sections, we evaluate $W_j$ and then determine $W$ and $p$ 
from Eq.\ (\ref{constr3}).

The numerator in $E_{\rm var}$ is evaluated in a similar way.  
Let us consider the expectation value of an operator $\widehat{\cal O}$ 
(such as $S_\ell^+ S_m^-$)
which operates on a certain part inside a cell of the $i_0$-th state.
(We call this cell as the central cell.)
Then all the configurations in the other cells are classified 
by $\{ N_i' \}$ which is the number of cells of the $i$-th state.
Then the expectation value becomes
\be
\langle \psi_0 | P_G \widehat{\cal O} P_G  | \psi_0\rangle 
= \sum_{i_0} \sum_{\{ N_i'\} } 
\frac{\bigl(\frac{N}{N_c}-1 \bigr)!}{\prod_{i=1}^K(N_i')!}
\prod_{i=1}^K \omega_i^{N_i'}
\langle \widehat{\cal O} \rangle_{i_0}, 
\label{Oexp}
\ee
in the analogy with Eq.\ (\ref{GAexp}).
Here $\langle \widehat{\cal O} \rangle_{i_0}$ indicates the expectation value 
of $\widehat{\cal O}$ together with terms such as 
$\prod_i \widehat n_{i\uparrow}(1-\widehat n_{i\downarrow})$ 
inside the central cell of the  $i_0$-th state. 

Determining the largest terms on the right-hand side and 
taking the ratio to the denominator, we finally obtain
\beq
\langle \widehat{\cal O} \rangle \equiv 
\frac{\langle \psi_0 | P_G \widehat{\cal O} P_G  | \psi_0\rangle }
{\langle \psi_0 | P_G P_G  | \psi_0\rangle } 
&&= \sum_{i_0}\frac{1}{\bigl(\frac{N}{N_c}\bigr)} N_{i_0} \omega_{i_0}^{-1}
\langle \widehat{\cal O} \rangle_{i_0} \nonumber\\
&&= \sum_{i_0}\frac{p^\nhiz}{W} \langle \widehat{\cal O} \rangle_{i_0}.
\label{numegen}
\eeq
The evaluation of the largest term and the derivation of the above 
result %(\ref{numegen}) 
is shown in Appendix A.

If we consider only the site-diagonal expectation values to 
estimate $\omega_i$, Eq.\ (\ref{numegen}) reproduces the original 
Gutzwiller approximation shown in the previous section.  
This is summarized in Appendix B.

\section{Half-Filled Case}

To obtain the Gutzwiller factors formulated in the 
last section, it is necessary to evaluate the weights $\omega_i$ 
for each state.  
Although these weights are generally complicated, 
we estimate them assuming that the corrections to the original 
Gutzwiller approximation (reproduced in Appendix B) are small.  
To be more specific, we evaluate $W_j$ in the lowest orders 
with respect to the intersite correlations 
$\xx$ and $\dd$ defined in Eq.\ (\ref{defxx}).
Typical values for $\xx$ and $\dd$ are less than $0.2$, so that the 
perturbation with respect to them will be justified.  

In this section we calculate the Gutzwiller approximation for the 
half filling, since it is less complicated than the doped case.  
All the cells do not contain holes 
so that only the weight, $W_0$, is to be calculated.  We call 
this subset of states as 0-hole sector.
After the half-filled case, it is rather straightforward to extend 
the results to the doped case.  

\subsection{Evaluation of $W_0$}
First we calculate the weight $\omega_i$ of the $i$-th state which 
has $n_{\uparrow i}$ up-spin electrons, $n_{\downarrow i}$ down-spin 
electrons:
\be
\omega_i = 
\langle 
\prod_{j\in {\cal A}} \widehat n_{j\uparrow}(1-\widehat n_{j\downarrow}) 
\prod_{j'\in {\cal B}} \widehat n_{j'\downarrow}(1-\widehat n_{j'\uparrow}) 
\rangle_0.
\ee
In the lowest order of $\xx$ and $\dd$ or in the zeroth order, we have to 
take only the site-diagonal expectation values.  This gives
\beq
\omega_i^{(0)}= [r(1-w)]^{n_{\rm right}} [w(1-r)]^{n_{\rm wrong}},
\eeq
where $n_{\rm right}$ ($n_{\rm wrong}$) means the number of sites
on which the right (wrong) spices of spin direction is located depending on 
sublattice 1 and 2.  The summation over all the possible 
configurations gives
\beq
W_0^{(0)} = [r(1-w)+w(1-r)]^{N_c} = (n-2rw)^{N_c}.
\eeq

In the next order with respect to $\xx$ and $\dd$, we have to consider 
contributions from expectation values of bonds in the cell as shown 
in Fig.\ 3.
For example, consider a bond connecting the 
sites $i$ and $j$ (see Fig.\ 3(a)) and a state in which two up-spin 
electrons occupy both $i$ and $j$ sites.  The expectation value in this 
bond gives
\beq
P_{\uparrow\uparrow} &&=
\langle (1-\widehat n_{i\downarrow}) \widehat n_{i\uparrow} 
\widehat n_{j\uparrow} (1-\widehat n_{j\downarrow}) \rangle_0 \nonumber\\
&&=rw(1-r)(1-w)-rw\xx^2 -(1-r)(1-w)\xx^2 \nonumber\\
&& -r(1-r)\dd^2 - w(1-w)\dd^2 +(\xx^2+\dd^2)^2,
\eeq
where we have used the Wick's theorem.  Generally $\xx$ and $\dd$ can 
be complex numbers and $\xx^2, \dd^2$ in the above expression 
mean $|\xx|^2, |\dd|^2$ implicitly.  In the similar way we calculate 
$P_{\uparrow\downarrow}, P_{\downarrow\uparrow}, P_{\downarrow\downarrow}$.
Writing that the left-spin in the subscript 
indicates the spin on the sublattice 1, they are calculated as
\beq
P_{\downarrow\downarrow} 
%&&=\langle (1-\widehat n_{i\uparrow}) \widehat n_{i\downarrow} 
%\widehat n_{j\downarrow} (1-\widehat n_{j\uparrow}) \rangle_0 
&&= P_{\uparrow\uparrow}, \nonumber\\
P_{\uparrow\downarrow} 
&&=  r^2(1-w)^2+2r(1-w)\xx^2                   \nonumber\\ 
&&\qquad\quad +r^2\dd^2 + (1-w)^2\dd^2 +(\xx^2+\dd^2)^2, \nonumber\\
P_{\downarrow\uparrow} 
&&=w^2(1-r)^2+2w(1-r)\xx^2                     \nonumber\\
&&\qquad\quad +w^2\dd^2 + (1-r)^2\dd^2 +(\xx^2+\dd^2)^2.
\eeq
\begin{figure}
\psfig{figure=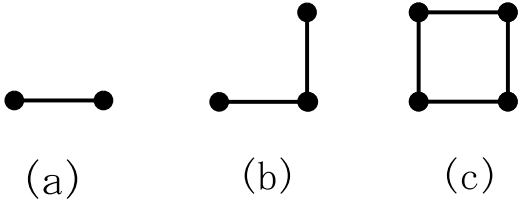,height=3cm}
\caption{Real-space diagrams for the evaluation of the weighting 
factors of the 0-hole cells.  
(a) Contribution from a bond which gives $X$ in $W_0^{(1)}$ and 
(b)(c) neglected diagrams. 
\label{fig:3}}
\end{figure}

In the estimation of $W_0^{(1)}$ we need their summation
\beq
P_{\uparrow\uparrow}+P_{\uparrow\downarrow}
+P_{\downarrow\uparrow}+P_{\downarrow\downarrow}
=(n-2rw)^2 + X,
\label{totalP}
\eeq
with $X$ defined in Eq.\ (\ref{XX2}), 
%\beq
%X&&=2\delta^2 (\dd^2-\xx^2)+8m^2(\xx^2+\dd^2)+4(\xx^2+\dd^2)^2, \nonumber\\
%&&\ 
%\eeq
where we have used the relations $r=n/2+m$ and $w=n/2-m$.  
The first term $(n-2rw)^2$ on the r.h.s.\ is the zeroth order 
contribution which has been included in $W_0^{(0)}$.  
In this sense, $X$ is a kind of connected (or cummulant) contribution 
coming from the real-space diagram in Fig.\ 3(a).
The actual values of $X$ is roughly $1/20$ so that our perturbation 
scheme can be justified.  
Denoting the number of bonds in a cell as $N_b$, 
we have a contribution to $W_0$ as
\beq
W_0^{(1)} = N_b X (n-2rw)^{N_c-2}, 
\label{w00}
\eeq
where the factor $(n-2rw)^{N_c-2}$ comes from the contributions 
of the other $N_c-2$ sites in the cell except for $i$ and $j$.  

In the similar way, we evaluate the higher order contributions to $W_0$. 
By considering two bonds in the cell 
we approximate their contribution as 
\beq
W_0^{(2)} = _{N_b}C_2 X^2 (n-2rw)^{N_c-4},
\label{secondc}
\eeq
where $_{N_b}C_2=N_b!/(N_b-2)!2!$ is the number of choices for the 
positions of the two bonds in a cell.  
The summation of these series leads to 
\beq
W_0 
&&= \sum_{j=0}^{N_b} \ _{N_b}C_j X^j (n-2rw)^{N_c-2j} \nonumber\\
&&= (n-2rw)^{N_c} \biggl( 1+\frac{X}{(n-2rw)^2}\biggr)^{N_b}.
\label{w0}
\eeq

Let us discuss the neglected contributions to $W_0$.  
Figures 3(b) and 3(c) show real-space diagrams whose connected expectation 
values can contribute to $W_0$.  
Although their contributions are not included, it can 
be shown that they are smaller than the terms in (\ref{w0}).
For example, the contribution from Fig.\ 3(b) is in the order of 
$N_b \delta^2 \xx^4$ and $N_b m^2 \xx^4$ etc., so that it is smaller 
than $W_0^{(1)}$.  
The contribution from Fig.\ 3(c) is in the order of $X^2 N_b$ 
which is smaller than (\ref{secondc}) by a factor $1/N_b$.  

There is another effect neglected in $W_0$: 
The higher order terms in 
(\ref{w0}) become less and less correct because the number of the bonds 
giving $X$ is not exactly $_{N_b}C_j$ due to their overlapping.  
However the error is again in the order smaller than $_{N_b}C_j$.  
In this sense, Eq.\ (\ref{w0}) is a kind of a summation 
of most dominant terms in the order of $X^j N_b^j$.

\subsection{Evaluation of $g_s^{XY}$}

In order to obtain the Gutzwiller factor for $S_\ell^+ S_m^-$ at
half filling we need to calculate 
\beq
\langle S_\ell^+ S_m^- \rangle 
=\sum_{i_0} \frac{1}{W_0} \langle S_\ell^+ S_m^- \rangle_{i_0},
\eeq
according to the general formulation (\ref{numegen}).
Note that $W=W_0$ and 
the central cell (which includes sites $\ell$ and $m$) does not 
contain holes, i.e., $\nhiz =0$.  
The average in a central cell 
of the $i_0$-th state, $\langle S_\ell^+ S_m^- \rangle_{i_0}$, is evaluated
in a similar perturbation scheme as in $W_0$.  
In the lowest order of $\xx$ and $\dd$, we have (Fig.\ 4(a))  
\beq
\langle S_\ell^+ S_m^- \rangle_0 (n-2rw)^{N_c-2}, 
\label{gsxy0}
\eeq
where the factor $(n-2rw)^{N_c-2}$ comes from the contributions 
from the sites in the central cell other than the sites $\ell$ 
and $m$.  
\begin{figure}
\psfig{figure=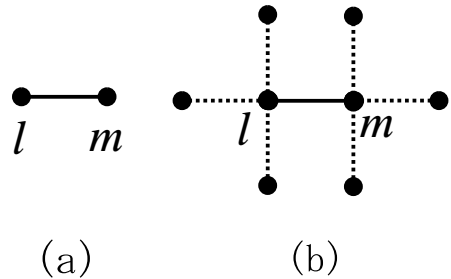,height=4cm}
\caption{Real-space diagrams for the evaluation of 
$\langle S_\ell^+ S_m^- \rangle$.  
(a) In the lowest order calculation, we have 
$\langle S_\ell^+ S_m^- \rangle_0$ for the sites $\ell$ and $m$.  
(b) In the next order, the contributions of $X$ come from the bonds
which are not connected to the sites $\ell$ and $m$.  
The dotted lines and the solid line are the excluded bonds.  
\label{fig:4}}
\end{figure}

\def\~#1{{\widetilde #1}}
\def\2~#1{{\widetilde #1}'}
\def\3~#1{{\widetilde #1}''}

In the next order of $\xx$ and $\dd$, there are contributions 
of $X$ from the bonds in the cell.  This leads to
\beq
\~{N_b} X \langle S_\ell^+ S_m^- \rangle_0 (n-2rw)^{N_c-4},
\label{gsxy1}
\eeq
where $\~{N_b}$ is the number of bonds which are not connected 
to the sites $\ell$ and $m$ directly (Fig.\ 4(b)).  
In general $\~{N_b}$ depends on the shape of the cell.  
However, if we choose the large enough cell, we have
\beq 
\~{N_b}=N_b-7,
\label{NB7}
\eeq
as is evident from Fig.\ 4(b).  

Although the dotted bonds in Fig.\ 4(b) do not give $X$, they may 
still give different contributions to 
$\langle S_\ell^+ S_m^- \rangle_{i_0}$, such as 
\beq
\langle S_\ell^+ S_m^- 
\widehat n_{m'\uparrow} (1-\widehat n_{m'\downarrow}) \rangle_c,
\label{gsxynn}
\eeq
where $m'$ is a nearest neighbor site of $m$ and 
$\langle \cdots \rangle_c$ means a connected expectation value 
excluding the disconnected terms such as 
$
\langle S_\ell^+ S_m^- \rangle_0
\langle \widehat n_{m'\uparrow} (1-\widehat n_{m'\downarrow}) \rangle_0.
$
(The disconnected expectation values have been 
already taken into account in Eq.\ (\ref{gsxy0}).)  
By calculating (\ref{gsxynn}), however, we find that there are no 
connected conributions:  
Actually the two electron operators in 
$S_\ell^+ = c_{\ell\uparrow}^\dagger c_{\ell\downarrow}$
have to make contractions with the electron operators on the site $m$ 
in the Wick's expansion, which gives just the disconnected 
expectation value.

By summing up the series of corrections such as (\ref{gsxy1}) we obtain
\beq
&&\sum_{i_0} \langle S_\ell^+ S_m^- \rangle_{i_0}   \nonumber\\
&&\ = \langle S_\ell^+ S_m^- \rangle_0 \{ (n-2rw)^{N_c-2} \nonumber\\
&&\qquad\qquad\qquad + \~{N_b} X (n-2rw)^{N_c-4} + \cdots \} \nonumber\\
&&\ = \langle S_\ell^+ S_m^- \rangle_0 (n-2rw)^{N_c-2} 
\biggl( 1+\frac{X}{(n-2rw)^2}\biggr)^{\~{N_b}}.
\eeq
Combining with $W_0$, the final result for the extended Gutzwiller 
approximation for $S_\ell^+ S_m^-$ becomes
\beq
\langle S_\ell^+ S_m^- \rangle 
= \frac{\langle S_\ell^+ S_m^- \rangle_0}{(n-2rw)^{2}} 
\biggl( 1+\frac{X}{(n-2rw)^2}\biggr)^{-(N_b-\~{N_b})}.
\eeq
Since the Gutzwiller factor is defined as the ratio between 
$\langle S_\ell^+ S_m^- \rangle$ and $\langle S_\ell^+ S_m^- \rangle_0$,
we have
\beq
g_s^{XY} &&= \frac{1}{(n-2rw)^{2}} 
\biggl( 1+\frac{X}{(n-2rw)^2}\biggr)^{-(N_b-\~{N_b})} \nonumber\\
&&=\frac{a^{-(N_b-\~{N_b})}}{(n-2rw)^{2}},
\eeq
where we have put
\beq
a \equiv  1+\frac{X}{(n-2rw)^2},
\eeq
which appears frequently in the following discussion.  
If we choose a large enough cell like $N_c >> 2$, we obtain
\beq
g_s^{XY}=\frac{a^{-7}}{(n-2rw)^2},
\label{gXYhf}
\eeq
according to Eq.\ (\ref{NB7}).
Physically the factor $a^{-7}$ with $a>1$ represents the exclusion 
effect of the bonds as shown in Fig.\ 4(b).

\subsection{Evaluation of $g_s^Z$}

The Gutzwiller factor for the $z$-component of 
${\mybf{S}}_\ell \cdot {\mybf{S}}_m$ is calculated in the similar way.  
The lowest orders with respect to $\xx$ and $\dd$ give the 
similar results to (\ref{gsxy0}) and (\ref{gsxy1}), in which 
$\langle S_\ell^+ S_m^- \rangle_0$ is replaced with 
$\langle S_\ell^z S_m^z \rangle_0$.
However, for $\langle S_\ell^z S_m^z \rangle_{i_0}$, 
an additional contribution appears from 
the diagram in Fig.\ 5(a) which did not give contributions in 
$\langle S_\ell^+ S_m^- \rangle_{i_0}$.
This is the reason why the anisotropy 
for the Gutzwiller factor between $xy$ component and $z$ component
appears, which is the most important feature of our 
extended Gutzwiller approximation.   

Assuming that the site $\ell$ and $m'$ in Fig.\ 5(a) are 
in the sublattice 1, we have a contribution 
\beq
&&\langle S_\ell^z S_m^z 
\bigl\{ \widehat n_{m'\uparrow}(1-\widehat n_{m'\downarrow})
      + \widehat n_{m'\downarrow}(1-\widehat n_{m'\uparrow}) \bigr\} 
\rangle_c \nonumber \\
&&=\langle S_\ell^z \rangle_0 
\langle S_m^z \widehat n_{m'\uparrow}(1-\widehat n_{m'\downarrow})
      + S_m^z \widehat n_{m'\downarrow}(1-\widehat n_{m'\uparrow}) 
\rangle_c \nonumber \\
&& = \frac{m}{2}(P_{\uparrow\uparrow}-P_{\uparrow\downarrow}
+P_{\downarrow\uparrow}-P_{\downarrow\downarrow}) \nonumber\\
&& = -m^2 X_2,
\eeq 
with $X_2$ defined in Eq.\ (\ref{XX2}).
Since the number of bonds connected to the site $m$ is 
\beq
N_2 \equiv \frac{N_b-\~{N_b}-1}{2},
\eeq
the contribution to $\sum_{i_0} \langle S_\ell^z S_m^z \rangle_{i_0}$ 
turns out to be
\beq
-2N_2 m^2 X_2 (n-2rw)^{N_c-3},
\eeq
where the contributions from the bonds connected to the site $\ell$ 
are also included.  
\begin{figure}
\psfig{figure=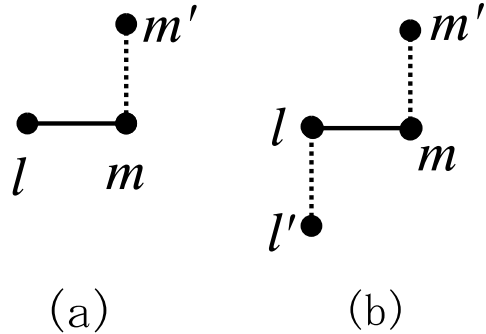,height=5cm}
\caption{Real-space diagrams contributing to 
$\langle S_\ell^z S_m^z \rangle$.  A kind of connected expectation 
values of (a) and (b) give additional contributions to $g_s^Z$,
compared with $g_s^{XY}$.  
\label{fig:5}}
\end{figure}

Similarly the diagram in Fig.\ 5(b) gives a contribution
\beq
&&\langle S_\ell^z \widehat n_{\ell'\uparrow}(1-\widehat n_{\ell'\downarrow})
      + S_\ell^z \widehat n_{\ell'\downarrow}(1-\widehat n_{\ell'\uparrow})
\rangle_c \nonumber\\
&&\times 
\langle S_m^z \widehat n_{m'\uparrow}(1-\widehat n_{m'\downarrow})
      + S_m^z \widehat n_{m'\downarrow}(1-\widehat n_{m'\uparrow}) 
\rangle_c \nonumber \\
&& = -\frac{1}{4}(P_{\uparrow\uparrow}-P_{\uparrow\downarrow}
+P_{\downarrow\uparrow}-P_{\downarrow\downarrow})^2 \nonumber\\
&& = -m^2 X_2^2.
\eeq 
By counting the number of possible combination of the bonds, we have
\beq
-N_2^2 m^2 X_2^2 (n-2rw)^{N_c-4}. 
\eeq

The higher order terms of $X^j$ are calculated as before, which become
\beq
\langle S_\ell^z S_m^z \rangle
= \frac{1}{W_0} \biggl\{ 
 &&\langle S_\ell^z S_m^z \rangle_0 (n-2rw)^{N_c-2} a^{\~{N_b}} \nonumber\\
-&&2N_2 m^2 X_2 (n-2rw)^{N_c-3} a^{\2~{N_b}} \nonumber\\
-&&N_2^2 m^2 X_2^2 (n-2rw)^{N_c-4} a^{\3~{N_b}} \biggr\} ,
\eeq
with $\2~{N_b}$ and $\3~{N_b}$ being the numbers of the bonds 
which are not connected directly to the diagrams in Figs.\ 5(a) and 5(b),
respectively.  They are shown in Figs.\ 6(a) and (b).
In the large enough cell, we have
\beq
\2~{N_b} &&= N_b -10, \nonumber\\
\3~{N_b} &&= N_b -13.
\eeq
Combining with $W_0$, we obtain
\beq
\langle S_\ell^z S_m^z \rangle
=&& \frac{a^{-(N_b-\~{N_b})}}{(n-2rw)^2} \biggl\{ 
 \langle S_\ell^z S_m^z \rangle_0 \nonumber\\
&&\qquad -\frac{2N_2}{n-2rw} m^2 X_2 a^{-(\~{N_b}-\2~{N_b})} \nonumber\\
&&\qquad -\frac{N_2^2}{(n-2rw)^2} m^2 X_2^2 a^{-(\~{N_b}-\3~{N_b})} \biggr\}.
\nonumber\\
&&\ 
\label{szsz}
\eeq
\begin{figure}
\psfig{figure=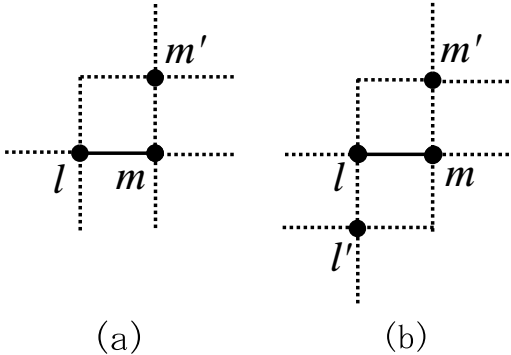,height=5cm}
\caption{The bonds excluded for the contribution of $X$, which 
are connected directly to $\ell, m$ and $m'$ sites in Fig.\ 5(a) 
and $\ell, m, m'$ and $\ell'$ sites in Fig.\ 5(b), respectively.  
\label{fig:6}}
\end{figure}

In this case, $\langle S_\ell^z S_m^z \rangle$ is not proportional to 
$\langle S_\ell^z S_m^z \rangle_0$.
Since the Gutzwiller factor is the ratio between the two, we have 
some nontrivial contribution from the second and third terms on the 
r.h.s of (\ref{szsz}).  Using 
\beq
\langle S_\ell^z S_m^z \rangle_0 = -m^2 - \frac{X_2}{4},
\eeq
the final expression for the Gutzwiller factor turns out to be
\beq
g_s^Z &&= g_s^{XY} \frac{1}{4m^2+X_2} \nonumber\\
&&\times \biggl[
X_2 + 4m^2 \bigl\{ 1+ \frac{N_2 X_2}{n-2rw} a^{-(\~{N_b}-\2~{N_b})}
\bigr\}^2 \biggr], 
\label{gsz}
\eeq
where we have assumed
\beq
\~{N_b} - \3~{N_b} = 2(\~{N_b} - \2~{N_b}).
\eeq
For a large enough cell we have $N_2=3$ and $\~{N_b} - \2~{N_b}=3$, 
so that 
\be
g_s^Z = g_s^{XY} \frac{1}{4m^2+X_2} \biggl[ 
X_2 + 4m^2 \bigl\{ 1+ \frac{3 X_2}{n-2rw} a^{-3} \bigr\}^2 \biggr].
\ee

The above formula is one of the most important results in this 
paper.  Typical $m$-dependences have been shown in Fig.\ 1. 
Let us here check some limiting cases.  
When $m=0$, we have
\be
g_s^{XY}=g_s^Z = 4 a^{-7},
\ee
which reproduces the original Gutzwiller approximation.  
For small $m$, we obtain
\be
g_s^Z = g_s^{XY} \left\{ 1+4m^2 \left( 
\frac{6}{n-2rw} a^{-3}+\frac{9 X_2}{(n-2rw)^2} a^{-6} \right)\right\}.
\ee
It is apparent that $g_s^Z$ has an enhancement as a function of $m$, 
which gives the reasonable AF long-range order at half filling.

\subsection{Physical meaning of the enhancement of $g_s^Z$}

In the usual interpretation of the Gutzwiller approximation, 
a comparison is made between the probabilities of spin configurations 
in the wave functions with and without the projection 
operators.\cite{ZGRS,Vollhardt}  
When we consider only two sites $\ell$ and $m$ for 
${\mybf{S}}_\ell \cdot {\mybf{S}}_m$, 
the summation of the probabilities of spin configurations 
$|\uparrow\uparrow \rangle, 
|\uparrow\downarrow \rangle, |\downarrow\uparrow \rangle$ and 
$|\downarrow\downarrow \rangle$ in the wave function {\it without} 
the projection is given by
$P_{\uparrow\uparrow}+P_{\uparrow\downarrow}
+P_{\downarrow\uparrow}+P_{\downarrow\downarrow}$
which was calculated as $(n-2rw)^2 +X$ in Eq.\ (\ref{totalP}).
On the other hand, in the wave function {\it with} the 
projection, we have always one of the above 
four states at half filling, and thus the probability is equal to 1.  
The ratio of these probabilities gives the Gutzwiller factor 
\beq
g_s^{XY} = g_s^Z = \frac{1}{(n-2rw)^2+X}.
\label{2sitegs}
\eeq
However this Gutzwiller approximation does not give a reasonable 
answer compared with the VMC results as mentioned before.

For the physical understanding of the results in the previous 
subsections, it is necessary 
to go beyond the two sites, $\ell$ and $m$, because the enhancement 
of $g_s^Z$ comes from the diagrams in Fig.\ 5.  
For this purpose we consider the spin configurations on three sites 
as shown in Fig.\ 7.  
To calculate the probabilities of these configurations 
is straightforward.  For example for Fig.\ 7(c) we have
\beq
\pmatrix{\ && \uparrow \cr \uparrow && \downarrow} = 
2 r(1-w) P_{\uparrow\downarrow} - 2r^3(1-w)^3,
\eeq
where the second term on the r.h.s.\ appears to avoid the 
double counting of the 
zeroth order with respect to $\xx$ and $\dd$.  
We can see that the presence of the third site $m'$ enhances the 
probability of this configuration,
$\pmatrix{\ && \uparrow \cr \uparrow && \downarrow}$, 
because $P_{\uparrow\downarrow}>P_{\uparrow\uparrow}
=P_{\downarrow\downarrow}>P_{\downarrow\uparrow}$ in the presence 
of AF order.
As a result, the weight of the configuration 
$\uparrow \downarrow$ on $\ell$ and $m$ sites is increased in the 
wave function.  This causes the enhancement of $g_s^Z$.  
\begin{figure}
\psfig{figure=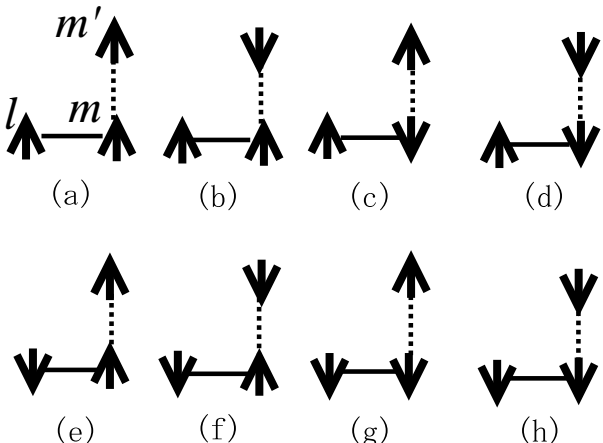,height=6cm}
\caption{All the spin configurations on $\ell, m$ and $m'$ sites, 
which contribute to $g_s^Z$.  
The probability of the configuration (c) is the largest because of 
the AF correlations ($\ell$ and $m'$ sites being on the sublattice 1).  
The effect of the surrounding AF correlations 
(in this case on the site $m'$)
is the physical origin of the enhancement of $g_s^Z$.  
\label{fig:7}}
\end{figure}

To understand this effect more quantitatively, we calculate 
$\langle S_\ell^z S_m^z \rangle$ directly from the configurations 
in Fig.\ 7 as follows:
\beq
&&4 \langle S_\ell^z S_m^z \rangle \nonumber\\
&&=\frac{{\rm (a)+(b)-(c)-(d)-(e)-(f)+(g)+(h)}}
{{\rm (a)+(b)+(c)+(d)+(e)+(f)+(g)+(h)}}.
\eeq
The denominator becomes
\beq
(n-2rw)\{ (n-2rw)^2 + 2X \},
\eeq
while the numerator becomes
\beq
&&(n-2rw)(P_{\uparrow\uparrow}-P_{\uparrow\downarrow}
-P_{\downarrow\uparrow}+P_{\downarrow\downarrow}) \nonumber\\
&&+2m(P_{\uparrow\uparrow}-P_{\uparrow\downarrow}
 +P_{\downarrow\uparrow}-P_{\downarrow\downarrow}) 
+4m^2(n-2rw) \nonumber\\
&&=4(n-2rw) \langle S_\ell^z S_m^z \rangle_0 -4m^2X_2.
\eeq
Although the denominator is larger than the two-site case in 
Eq.\ (\ref{2sitegs}), 
the second term in the numerator enhances 
$\langle S_\ell^z S_m^z \rangle $, which is mainly from the contribution 
of the configuration 7(c).  
Using $4\langle S_\ell^z S_m^z \rangle_0 =-(4m^2+X_2)$ and 
taking the ratio, we obtain
\beq
\frac{\langle S_\ell^z S_m^z \rangle}{\langle S_\ell^z S_m^z \rangle_0}
&&=\frac{1}{(n-2rw)^2+2X}\frac{1}{4m^2+X_2} \nonumber\\
&&\times \biggl\{ X_2 + 4m^2 (1+\frac{X_2}{n-2rw} ) \biggr\}.
\eeq
This is essentially the result obtained in the previous 
subsection for $g_s^Z$.  
Since we have considered only the site $m'$ as the third 
term, $N_2$ in Eq.\ (\ref{gsz}) is replaced with $1$ in this case.  
The first factor $1/\{(n-2rw)^2+2X\}$ corresponds the 
exclusion effect which was represented by $a^{-(N_b-\~{N_b})}$ 
in $g_s^{XY}$.  
From this analysis it is apparent 
that the enhancement of $g_s^Z$ as a function of $m$
is due to the increase of the probability of $\uparrow\downarrow$ 
in the presence of the AF circumstances.

\section{Less-than-half-filled case}

\subsection{Evaluation of the total weight $W$}
In the presence of holes, the Gutzwiller factors are calculated 
similarly as in the half-filled case.  
In this subsection we obtain the total weight, $W$.  
Since the weight for the zero-hole sector $W_0$ has been calculated, 
we have to evaluate 
\beq
W_1 = \sum_{i\ {\rm with} \ 1 \ {\rm hole}} \omega_i,
\eeq
and $W_2$ and so on.

In the lowest order of $\xx$ and $\dd$, we have
\beq
W_1^{(0)} = N_c (1-r)(1-w) (n-2rw)^{N_c-1}, 
\eeq
where the factor $N_c$ comes from the possible position of the hole 
in a cell and 
the factor $(1-r)(1-w)$ comes from the expectation value of the hole 
position, 
$\langle (1-\widehat n_{i\uparrow})(1-\widehat n_{i\downarrow})\rangle_0$.
In the next order, by counting the number of bonds which contribute 
$X$, we obtain
\beq
N_c (1-r)(1-w) N_{1b} X (n-2rw)^{N_c-3}, 
\eeq
where $N_{1b}$ means the number of bonds which are not connected 
the hole site as shown in Fig.\ 8(a).  
In the large enough cell we have
\beq
N_{1b}= N_b -4.
\eeq
\begin{figure}
\psfig{figure=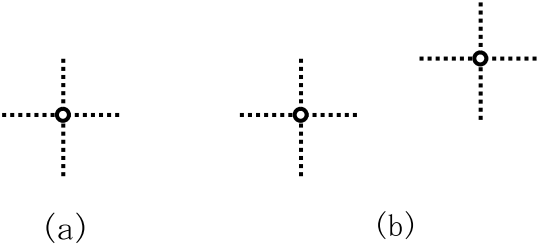,height=3.5cm}
\caption{The bonds excluded for the contribution of $X$, which are 
connected directly to (a) the hole site, and (b) the sites of two holes.  
\label{fig:8}}
\end{figure}

Although the dotted bonds in Fig.\ 8(a) do not give $X$, they give 
alternative contributions to $W_1$ in the same order.  
This contribution is calculated as
\beq
&&\langle (1-\widehat n_{i\uparrow})(1-\widehat n_{i\downarrow})
\{ \widehat n_{j\uparrow}(1-\widehat n_{j\downarrow})
 + \widehat n_{j\downarrow}(1-\widehat n_{j\uparrow}) \}
\rangle_0 \nonumber\\
&&=(n-2rw)(1-r)(1-w)+Y,
\eeq
with
\be
Y=\delta (1+\delta)(\xx^2-\dd^2) -4m^2(\xx^2+\dd^2) -2 (\xx^2+\dd^2)^2.
%Y&&=\delta (1+\delta)(\xx^2-\dd^2) -4m^2(\xx^2+\dd^2) \nonumber\\
%&&-2 (\xx^2+\dd^2)^2.
\ee
Here $i$ represents the site of the hole and $j$ is one of the 
nearest neighbor sites.  Since the number of the dotted bonds 
in Fig.\ 8(a) is $(N_b-N_{1b})$, the additional contribution to 
$W_1$ becomes
\beq
N_c (N_b-N_{1b}) Y (n-2rw)^{N_c-2}.
\eeq

The higher order terms with respect to $X$ can be taken into 
account in the similar way as before and we obtain
\beq
W_1&&= N_c (1-r)(1-w) (n-2rw)^{N_c-1} a^{N_{1b}}   \nonumber\\
  && + N_c (N_b-N_{1b}) Y (n-2rw)^{N_c-2} a^{\~{N_b}},
\eeq
where $\~{N_b}$ appears because the diagram in Fig.\ 8(a) excludes the 
same number of bonds as in Fig.\ 5(a).  We assume that 
\beq
\~{N_b}=N_{1b}-3,
\eeq
which leads to 
\beq
W_1&&= N_c \left\{ (1-r)(1-w)+\frac{N_b-N_{1b}}{n-2rw} Ya^{-3} \right\}
\nonumber\\
&&\times (n-2rw)^{N_c-1}a^{N_{1b}}.  
\eeq
Strictly speaking, when the hole is located on the boundary of 
the cell, the weight should be different.  
However we make an approximation of large enough cell.  

Continuing the similar arguments, we obtain
\beq
W_2&&=_{N_c}C_2 
\biggl\{ (1-r)(1-w)+\frac{N_b-N_{1b}}{n-2rw} Ya^{-3} \biggr\}^2 \nonumber\\
&&\times (n-2rw)^{N_c-2}a^{N_{2b}}, 
\label{W2lthf}
\eeq
for the two-hole sector, 
where $N_{2b}$ is the number of bonds which are not connected to 
the two sites of holes (Fig.\ 8(b)).  We have neglected the 
case in which two holes come to the nearest neighbor sites.
From Eq.\ (\ref{W2lthf}) it is apparent that 
the $j$-hole sector, $W_j$, can be generally approximated as 
\beq
W_j =_{N_c}C_j z^j (n-2rw)^{N_c-j} a^{N_{j,b}},
\eeq
with
\beq
z=(1-r)(1-w)+\frac{N_b-N_{1b}}{n-2rw} Ya^{-3}.  
\eeq

Finally approximating that $N_{j,b}=N_b-4j$, the constraints which 
determine $W$ and $p$ (Eq.\ (\ref{constr3})) become
\beq
\sum_{j=0}^{N_c} \frac{W_j}{W}p^j 
&&= \frac{1}{W}(n-2rw+pza^{-4})^{N_c}a^{N_b}=1, \nonumber\\
\sum_{j=0}^{N_c} j\frac{W_j}{W}p^j 
&&= \frac{N_c}{W}pza^{-4}(n-2rw+pza^{-4})^{N_c-1}a^{N_b} \nonumber\\
&&=\MYdelta N_c.
\eeq
From these constraints we obtain
\beq
p &&= \frac{\delta(n-2rw)}{nz}a^{4}, \nonumber\\
W &&= \biggl( \frac{n-2rw}{n}\biggr)^{N_c} a^{N_b},
\label{pandW}
\eeq
for less-than-half-filling.

\subsection{Evaluation of $g_s^{XY}$ and $g_s^Z$}

In the similar way we estimate $g_s^{XY}$.  In the case 
when there are $j$ holes in the central cell, we have
\beq
&&\sum_{i_0\ {\rm with}\ j\ {\rm holes}} 
\langle S_\ell^+ S_m^- \rangle_{i_0} \nonumber\\
&&=_{N_c-2}C_j z^j (n-2rw)^{N_c-2-j}
a^{\~{N_b}-4j} \langle S_\ell^+ S_m^- \rangle_{0}.
\eeq
By summing up all the sectors with different number of holes, we have
(according to the general formula (\ref{numegen})),
\beq
\langle S_\ell^+ S_m^- \rangle &&= \sum_{j=0}^{N_c-2} \frac{p^j}{W} 
\sum_{i_0\ {\rm with}\ j\ {\rm holes}} 
\langle S_\ell^+ S_m^- \rangle_{i_0} \nonumber\\
&&=\frac{1}{W}(n-2rw+pza^{-4})^{N_c-2}a^{\~{N_b}}
\langle S_\ell^+ S_m^- \rangle_{0} \nonumber\\
&&=\biggl(\frac{n}{n-2rw}\biggr)^2 a^{-(N_b-\~{N_b})}
\langle S_\ell^+ S_m^- \rangle_{0}.
\eeq
From this we obtain the Gutzwiller factor 
\beq
g_s^{XY}=\biggl(\frac{n}{n-2rw}\biggr)^2 a^{-7}.
\eeq
This is a simple generalization of the results at half filling 
obtained in Eq.\ (\ref{gXYhf}). 

Although the estimation of $g_s^Z$ is essentially the same as in the 
half-filled case, there are some additional diagrams to be 
taken into account which are shown in Fig.\ 9.  
The detailed calculations are summarized in Appendix C.  The final 
result is 
\beq
g_s^Z &&= g_s^{XY} \frac{1}{4m^2+X_2} \nonumber\\
&&\times \biggl[ X_2 + 4m^2 \bigl\{
1+\frac{N_2 X_2 n}{n-2rw} \bigl(1-\frac{p}{2}\bigr) a^{-3}\bigr\}^2 \biggr].
\eeq
The quantity $p$ is given in Eq.\ (\ref{pandW}).  For $g_s^Z$, 
we use the lowest order approximation 
\be
p \sim \frac{\delta(n-2rw)}{nz} \sim 2\delta,
\ee
since $n-2rw=1/2$ and $z=1/4$ in the lowest order with respect to 
$\delta, \xx$ and $\dd$.  
This gives our final expression for $g_s^Z$ given in 
Eq.\ (\ref{gszfinal}).

\begin{figure}
\psfig{figure=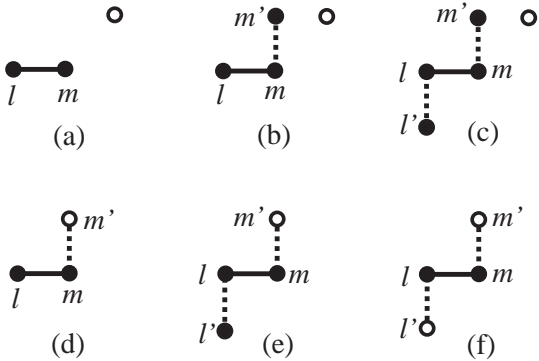,height=5cm}
\caption{Real-space diagrams which give additional contributions 
to $g_s^Z$ in the presence of holes.  The detailed calculation 
for each configuration is summarized in Appendix C.
\label{fig:9}}
\end{figure}

\subsection{Evaluation of $g_t$}
As for the Gutzwiller factor for the hopping term, $g_t$, 
we have to calculate 
\beq
\langle (1-\widehat n_{i\downarrow}) c_{i\uparrow}^\dagger 
c_{j\uparrow}(1-\widehat n_{j\downarrow})
\rangle_{i_0}.
\eeq
Since the hopping term needs at least one hole, the lowest sector is 
the one-hole sector.  For this sector, we have
\beq
&&\sum_{i_0\ {\rm with}\ 1\ {\rm hole}}
\langle (1-\widehat n_{i\downarrow}) c_{i\uparrow}^\dagger 
c_{j\uparrow}(1-\widehat n_{j\downarrow})
\rangle_{i_0} \nonumber\\
&&= \langle (1-\widehat n_{i\downarrow}) c_{i\uparrow}^\dagger 
c_{j\uparrow}(1-\widehat n_{j\downarrow})
\rangle_{0} a^{\~{N_b}} (n-2rw)^{N_c-2} \nonumber\\
&&= T a^{\~{N_b}} (n-2rw)^{N_c-2},
\eeq
where the expectation value of the correlated hopping, $T$, is 
calculated as
\beq
T &&\equiv \langle (1-\widehat n_{i\downarrow}) c_{i\uparrow}^\dagger 
c_{j\uparrow}(1-\widehat n_{j\downarrow}) \rangle_{0} \nonumber\\
&&= (1-r)(1-w)\xx - \xx^3 -\xx\dd^2 \nonumber\\
&&= \xx\{ (1-r)(1-w) - \frac{X_2}{2} \}.
\eeq
It is easy to show that for the $j+1$-hole sector 
\beq
\ _{N_c-2}C_j  T a^{\~{N_{j,b}}} (n-2rw)^{N_c-2-j} z^j.
\eeq
Therefore summing up all the sectors, we obtain
\beq
\langle c_{i\uparrow}^\dagger c_{j\uparrow} \rangle 
&&= \frac{1}{W} \sum_{j=0}^{N_c-2}
\ _{N_c-2}C_j  T a^{\~{N_{j,b}}} (n-2rw)^{N_c-2-j} z^j p^{j+1} \nonumber\\
&&= \biggl(\frac{n}{n-2rw}\biggr)^2 T a^{-(N_b-\~{N_b})} p.
\eeq
Since $\langle c_{i\uparrow}^\dagger c_{j\uparrow} \rangle_0 = \xx$, 
the Gutzwiller factor 
$g_t=\langle c_{i\uparrow}^\dagger c_{j\uparrow} \rangle/
\langle c_{i\uparrow}^\dagger c_{j\uparrow} \rangle_0$ becomes
\beq
g_t &&= \biggl(\frac{n}{n-2rw}\biggr)^2 
\{ (1-r)(1-w) - \frac{X_2}{2} \} a^{-(N_b-\~{N_b})} p \nonumber\\
&&= \frac{n}{n-2rw}
\frac{\delta \{ (1-r)(1-w) - \frac{X_2}{2} \}}{z} 
a^{-(N_b-\~{N_b})+4}, \nonumber\\
&&\ 
\eeq
where we have substituted the value of $p$ obtained in Eq.\ (\ref{pandW}).
The diagrams in Fig.\ 5 give contributions of the order of 
$\delta^2\xx^2$ or $\delta m^2 \xx^2$ so that they are neglected.

Finally comparing the definitions of $X$ and $Y$, we use an 
approximation 
\beq
Y\sim -\frac{X}{2},
\eeq
for the small $\delta$ case.  
Using this approximation, the quantity $z$ can be rewritten as
\beq
z 
&&=(1-r)(1-w) 
\biggl\{ 1+\frac{N_b-N_{1b}}{(1-r)(1-w)(n-2rw)} Ya^{-3} \biggr\} \nonumber\\
&&=(1-r)(1-w) \bigl\{ 1-4(N_b-N_{1b})X \bigr\} \nonumber\\
&&=(1-r)(1-w) a^{-(N_b-N_{1b})} \nonumber\\
&&=(1-r)(1-w) a^{-4}.
\eeq
Here we are considering $\delta, m$ and $\xx$ as small quantities.  
Substituting this approximate $z$ into $g_t$ we have
\beq
g_t &&= \frac{\delta n}{n-2rw}
\frac{(1-r)(1-w) - \frac{X_2}{2} }{(1-r)(1-w)} a^{-(N_b-\~{N_b})+8} 
\nonumber \\
&&=\frac{2\delta (1-\delta)}{1-\delta^2+4m^2} \ 
\frac{(1+\delta)^2-4m^2-2X_2}{(1+\delta)^2-4m^2}a.
\eeq
This is our final expression for $g_t$ given in Eq.\ (\ref{gtfinal}).

\section{Optimized variational state}

In this section we calculate the energy of the variational state  
using the Gutzwiller factors obtained in the previous sections.  
We will show that the variational energies and the magnitudes of 
order parameters agree with the results in VMC simulations.  

In the variational state 
$|\psi\rangle=P_G|\psi_0(\Delta_d^{\rm V}, \Delta_{\rm af}^{\rm V}, 
\mu^{\rm V})\rangle$ in Eq.\ (\ref{vwf}), 
$|\psi_0(\Delta_d^{\rm V}, \Delta_{\rm af}^{\rm V}, \mu^{\rm V})\rangle$ 
is a Hartree-Fock type wave function 
with the $d$-wave SC and AF orders.
It is expressed as\cite{Himeda}
\begin{eqnarray}
&&|\psi_0(\Delta_d^{\rm V}, \Delta_{\rm af}^{\rm V}, \mu^{\rm V})\rangle
=\prod_{k,s(=\pm)}(u_k^{(s)}+v_k^{(s)}d^{(s)\dagger}_{k\uparrow}
    d^{(s)\dagger}_{-k\downarrow})|0\rangle \nonumber\\
&&\qquad\qquad =\prod_{k,s}u_k^{(s)}\exp\left[\sum_{k,s}
     \frac{v_k^{(s)}}{u_k^{(s)}}d^{(s)\dagger}_{k\uparrow}
    d^{(s)\dagger}_{-k\downarrow}
\right]|0\rangle,
\end{eqnarray}
where
\begin{eqnarray}
\frac{v_{k}^{(\pm)}}{u_{k}^{(\pm)}}
&=&\frac{\pm\Delta_d^{\rm V}\eta_k}
 {\left(\pm E_k-\mu^{\rm V}\right)
   +\sqrt{\left(\pm E_k-\mu^{\rm V}\right)^2
            +(\Delta_d^{\rm V}\eta_k)^2}}, \\
E_k&=&\sqrt{\epsilon_k^2+\Delta_{\rm af}^{{\rm V}2}},
\end{eqnarray}
$\epsilon_k=-t\gamma_k$, 
$\gamma_k=2(\cos{k_x}+\cos{k_y})$ and $\eta_k=2(\cos{k_x}-\cos{k_y})$.
The annihilation operators $d_{k\sigma}^{(s)}$ are related to the 
electron operators through the following unitary transformation,
\begin{equation}
\left(
   \begin{array}{c}
   d_{k\sigma}^{(+)} \\ 
   d_{k\sigma}^{(-)}
      \end{array}
\right) =
\left(\begin{array}{cc}
   \alpha_{k\sigma} & -\beta_{k\sigma} \\ \beta_{k\sigma} & \alpha_{k\sigma}
     \end{array}\right)
\left(\begin{array}{c}
     c_{Ak\sigma} \\ c_{Bk\sigma}
      \end{array}		
\right),
\end{equation}
with
\begin{equation}
\left\{ \begin{array}{c}
\alpha_{k\sigma}=\sqrt{\frac{1}{2}\left(1-\frac{\sigma\Delta_{\rm af}^{\rm V}}
              {E_k}\right)} \\
\beta_{k\sigma}=\sqrt{\frac{1}{2}\left(1+\frac{\sigma\Delta_{\rm af}^{\rm V}}
              {E_k}\right)} 
       \end{array}
\right. 
.
\end{equation}
Here $c_{Ak\sigma}(c_{Bk\sigma})$ are annihilation operators of
an electron on the A(B)-sublattice
and $\sigma$ represent $\uparrow$(+1) and $\downarrow$(-1).
The wave vector $\mybf{k}$ is limited to half of the Brillouin zone 
where $\epsilon_k<0$.
We can confirm that $|\psi _0\rangle$ is a vacuum 
of the annihilation operators which diagonalize
\begin{eqnarray}
 \sum_{k} && \Big[\sum_{\sigma} 
    \big\{\epsilon_k(c_{Ak\sigma}^{\dagger}c_{Bk\sigma}+h.c.)
         \nonumber \\
  &&-(\mu^{\rm V} +\sigma\Delta_{\rm af}^{\rm V})
       c_{Ak\sigma}^{\dagger}c_{Ak\sigma}
    -(\mu^{\rm V} -\sigma\Delta_{\rm af}^{\rm V})
       c_{Bk\sigma}^{\dagger}c_{Bk\sigma} \big\}    \nonumber \\
  &&- \Delta_d^{\rm V}\eta_k(c_{A-k\downarrow}c_{Bk\uparrow}
                   +c_{B-k\downarrow}c_{Ak\uparrow}+h.c.) \Big].
\label{MFH}
\end{eqnarray}
%Thus, this wave function is a natural extension of the coexistent 
%state between the $d$-wave SC and the AF orders.

In order to clarify the correspondence to the mean-field theory, 
let us consider the effective Hamiltonian, 
$\widehat{\cal H}_{\rm eff}$ in Eq.\ (\ref{Evar}).  
In $\widehat{\cal H}_{\rm eff}$ the parameter $t$ in $\cal H$ is 
replaced with 
\be
t_{\rm eff} = g_t t, 
\ee
and the exchange term is replaced with 
\beq
J{\mybf{S}}_i \cdot {\mybf{S}}_j 
= g_s^{XY} J \left( S^x_i S^x_j + S^y_i S^y_j \right)
+ g_s^Z    J S^z_i S^z_j ,  
\eeq
where $g_t$, $g_s^{XY}$ and $g_s^Z$ are the Gutzwiller factors 
obtained in the previous sections.  
When we apply the mean-field theory to $\widehat{\cal H}_{\rm eff}$, 
we obtain the similar Hamiltonian as in Eq.\ (\ref{MFH}) 
but with the replacements 
\beq
t\ \ &&\rightarrow t_{\rm eff} + J_{\rm eff} \xx^{\rm V}, \nonumber\\
\Delta_d^{\rm V} &&\rightarrow J_{\rm eff} \Delta^{\rm V}, \nonumber\\
\Delta_{\rm af}^{\rm V} &&\rightarrow 2J_{\rm eff}^Z m^{\rm V}, 
\eeq
with 
\be
J_{\rm eff} = \frac{1}{2} g_s^{XY} J + \frac{1}{4} g_s^Z J,
\ee
\be
J_{\rm eff}^Z = g_s^Z J.
\ee

Note that in the usual mean-field theory the self-consistency 
equations give $\xx^{\rm V}=\xx, \Delta^{\rm V}=\Delta$ and 
$m^{\rm V}=m$, where $\xx, \Delta$ and $m$ are the expectation values 
$\dd=\langle c_{i\uparrow}^{\dagger}c_{j\downarrow}^{\dagger}\rangle_0$,
$\xx=\langle c_{i\sigma}^{\dagger}c_{j\sigma}\rangle_0$, and 
$m=\frac{1}{2}(-1)^i \left( \langle\widehat{n_{i\uparrow}}\rangle_0 -
\langle\widehat{n_{i\downarrow}}\rangle_0 \right)$
used in the previous sections.  
We will see shortly that $\xx^{\rm V}$ and $\xx$ 
etc.\ are slightly different due to the dependence of the Gutzwiller 
factors on $\xx, \Delta$ and $m$.  

Using the wave function $|\psi_0\rangle$, the expectation values become
\begin{eqnarray}
\dd &=&\frac{1}{8N}\sum_{k,\pm}\frac{F_k \eta_k}
                 {\sqrt{(\pm E_k-\mu^{\rm V})^2+F_k^2}},\\
\xx &=&\frac{1}{8N}\sum_{k,\pm}\frac{(t_{\rm eff}+J_{\rm eff}\xx^{\rm V})
       \gamma_k^2 (\pm E_k-\mu^{\rm V})}
         {\pm E_k\sqrt{(\pm E_k-\mu^{\rm V})^2+F_k^2}}, \\
m &=&\frac{1}{2N}\sum_{k,\pm}\frac{2J_{\rm eff}^Z m^{\rm V} 
(\pm E_k-\mu^{\rm V} )}
         {\pm E_k\sqrt{(\pm E_k-\mu^{\rm V})^2+F_k^2}}, \\
n &=&1-\frac{1}{N}\sum_{k,\pm}\frac{(\pm E_k-\mu^{\rm V})}
                {\sqrt{(\pm E_k-\mu^{\rm V})^2+F_k^2}},
\end{eqnarray}
with
\beq
F_k &=& J_{\rm eff}\Delta^{\rm V} \eta_k, \\
E_k &=& \sqrt{
(t_{\rm eff}+J_{\rm eff}\xx^{\rm V})^2 \gamma_k^2 + 
(2J_{\rm eff}^Z m^{\rm V})^2},
\eeq
where $N$ is the number of sites and the summation over 
$\mybf{k}$ is limited to half of the Brillouin zone.   

Using these expectation values we obtain 
\beq
E_{\rm var} &=& \langle \widehat{\cal H}_{\rm eff} \rangle_0 \nonumber\\
&=& -8N t_{\rm eff} \xx_-4N J_{\rm eff} \left( \dd^2 + \xx^2 \right) 
\nonumber\\
& & -2N J_{\rm eff}^Z m^2 - \mu (Nn - N_{\rm e}).  
\eeq
By taking the derivatives with respect to the variational parameters, 
$\dd^{\rm V}, \xx^{\rm V}$, and $m^{\rm V}$, we find the following 
self-consistency equations:
\beq
\dd^{\rm V} &=& \dd - \frac{1}{8N J_{\rm eff}}
\langle \frac{\partial H_{\rm eff}}{\partial \dd} \rangle_0 ,\nonumber \\
\xx^{\rm V} &=& \xx - \frac{1}{8N J_{\rm eff}}
\langle \frac{\partial H_{\rm eff}}{\partial \xx} \rangle_0 ,\nonumber \\
m^{\rm V} &=& m - \frac{1}{4N J_{\rm eff}^Z}
\langle \frac{\partial H_{\rm eff}}{\partial m} \rangle_0 ,\nonumber \\
\mu^{\rm V} &=& \mu - \frac{1}{N}
\langle \frac{\partial H_{\rm eff}}{\partial n} \rangle_0,
\label{SCeq}
\eeq
where the partial derivative of $H_{\rm eff}$ is applied to the 
Gutzwiller factors, $g_s^{XY}, g_s^Z$ and $g_t$.  

Figure 10 shows the self-consistent parameters $\dd, \xx$ and $m$ 
satisfying (\ref{SCeq}) 
as a function of the doping $\delta=1-n$ for $J/t=0.3$.  
Note that these are the expectation values in the wave function 
$|\psi _0\rangle$ without the projection.  
The dashed line in Fig.\ 10 represents the results when the AF order is 
suppressed, i.e., $m$ is fixed to zero.  It is apparent that the 
presence of the AF order has a small effect on the expectation values 
$\dd$ and $\xx$.  
The coexistent state between the d-wave SC and AF order parameters 
is stabilize up to the doping rate $\delta=0.1$.  This is 
consistent with the VMC simulations.  
\begin{figure}
\psfig{figure=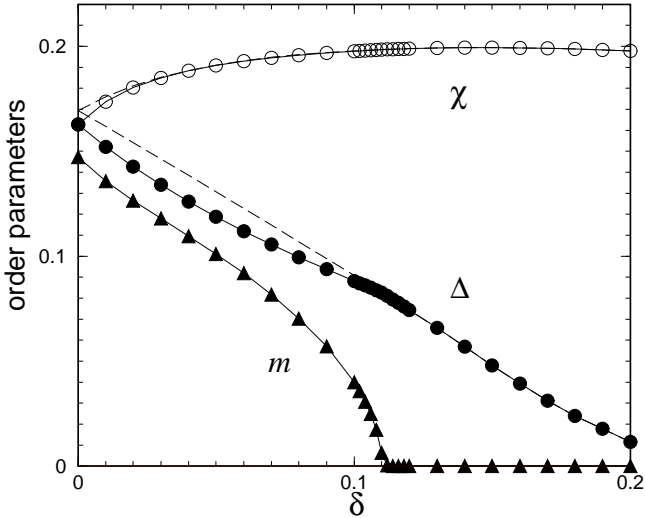,height=7cm}
\caption{The self-consistent parameters $\dd, \xx$ and $m$ 
as a function of the doping rate $\delta=1-n$ for $J/t=0.3$.  
The dashed line represents the results when the AF order is 
suppressed i.e, $m$ is fixed to zero.  
\label{fig:10}}
\end{figure}

The actual expectation values in the wave function $P_G |\psi_0\rangle$ 
with projection are different from $\dd, \xx$ and $m$.  
For example, the expectation value of $c_{i\sigma}^\dagger c_{j\sigma}$ is
\be
\xx_{\rm exp}
\equiv \langle c_{i\sigma}^\dagger c_{j\sigma}\rangle 
=g_t \langle c_{i\sigma}^\dagger c_{j\sigma}\rangle_0
=g_t \xx.
\ee
In the similar way, the actual expectation values in $P_G |\psi_0\rangle$ 
are obtained using corresponding Gutzwiller factors.  
For $m_{\rm exp}$ we repeat the similar arguments as for $g_s^Z$ to obtain 
\beq
&&m_{\rm exp} 
\equiv \frac{1}{2}\langle n_{i\uparrow} - n_{i\downarrow} \rangle 
= g_m m, \nonumber \\
&&g_m = \frac{n}{n-2rw} a^{-4}
\biggl[ 1+\frac{(N_b-N_{1b}) X_2 n}{n-2rw} \bigl(1-\frac{p}{2}\bigr) 
a^{-3} \biggr]. \nonumber\\
&&\ 
\eeq
For the expectation value of $c_{i\uparrow}^\dagger c_{j\downarrow}^\dagger$, 
we have to take care 
of the fact that the number of holes changes in the cell.  
Here we use the average number of holes as an approximation.  
In the same level of approximations as for $g_t$ we obtain
\beq
\dd_{\rm exp} &&\equiv \langle c_{i\uparrow}^\dagger c_{j\downarrow}^\dagger
\rangle = g_\Delta \dd, \nonumber \\
g_\Delta      &&= \frac{\delta n}{n-2rw}
\frac{(1-r)^2 + (1-w)^2 + X_2 }{2(1-r)(1-w)} a^{-(N_b-\~{N_b})+8} 
\nonumber\\
&&=\frac{2\delta (1-\delta)}{1-\delta^2+4m^2} \ 
\frac{(1+\delta)^2 + 4m^2 + 2X_2}{(1+\delta)^2-4m^2}a.
\eeq

Figure 11 shows the actual expectation values, $\dd_{\rm exp}$ 
and $m_{\rm exp}$ as a function of the doping $\delta=1-n$ for $J/t=0.3$.  
\begin{figure}
\psfig{figure=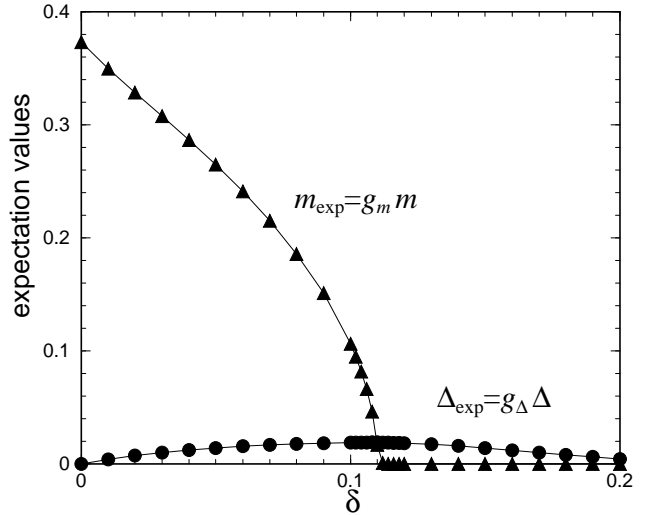,height=7cm}
\caption{The actual expectation values, $\dd_{\rm exp}$ 
and $m_{\rm exp}$ in the optimized (or self-consistent) wave function 
with the projection operator 
as a function of the doping rate $\delta=1-n$ for $J/t=0.3$.  
\label{fig:11}}
\end{figure}

\section{Summary and Discussion}

In this paper we have developed a new approach for studying the effect 
of strong correlation or the Gutzwiller's projection in 
the two-dimensional $t$-$J$ model.  It is based on the extended 
Gutzwiller approximation, in which the effects of longer-range correlations 
are taken into account.  
These correlations play important roles for 
the interplay between the AF and d-wave SC.  
Let us summarize our main results and discuss related problems.  

(1) Generally the expectation values with respect to the projected 
wave function are strongly renormalized due to the 
exclusion of double occupancies.  
In the slave-boson mean-field theory, this effect is taken into 
account by assuming the replacement $t\rightarrow \delta t$ and 
$J \rightarrow J$.  However we have shown that the renormalization 
is not so simple.  
First of all the renormalization factor for the exchange interaction 
is anisotropic, i.e., 
$J^{XY}_{\rm eff}=g_s^{XY} J \ne J^{Z}_{\rm eff}=g_s^{Z}J$ 
in the presence of AF moment.  
Furthermore $g_t, g_s^{XY}$ and $g_s^Z$ have nonlinear dependences 
on the expectation values $\dd, \xx$ and $m$.  
In this sense the extended Gutzwiller approximation is beyond the 
simple slave-boson mean-field theory. 
We think that essence of the strong correlation is contained in 
these Gutzwiller factors, because they stem solely from the projection.  

The physical meanings of the Gutzwiller factors are clarified:

(i) The factor $a^{-7}$ appearing in $g_s^{XY}$ and $g_s^Z$ represents 
the exclusion effect shown in Fig.\ 4(b).  When we calculate 
$\langle {\mybf{S}}_i {\mybf{S}}_j\rangle$, the surrounding six bonds 
cannot make contribution of nearest neighbor correlation $X$.  
As is evident in Fig.\ 1, the factor $a^{-7}$ reduces the value 
of $J_{\rm eff}$ even for the case $m=0$.  

(ii) The enhancement of $g_s^Z$ as a function of $m$ is the most 
important feature.  This was observed in the VMC simulations.\cite{Himeda} 
Here we have identified its origin 
as the effect of surrounding AF correlations by discussing the 
probabilities of spin configurations in Fig.\ 7.  
The enhancement of $g_s^Z$ is due to the increase of 
$\uparrow \downarrow$ configuration caused by the AF circumstances.  

(2) We have studied the projected variational states as an 
application of the new Gutzwiller approximation.  
It is shown that our scheme reproduces the results in the VMC 
simulations.\cite{Himeda}
The enhancement of $g_s^Z$ as a function of $m$ is essential for 
reproducing the coexistent state of the AF and d-wave SC orders 
found in VMC.\cite{Chen,GL,Himeda} 

(3) At half filling ($\delta=0$), there are several interesting 
features.  
In the original Gutzwiller approximation, the AF state at 
half filling was not obtained\cite{ZGRS} which was unphysical.  
However in the present Gutzwiller approximation, the AF state 
with d-wave SC order parameter becomes the most optimized state 
as obtained in VMC. 
Note that the expectation value of the SC order parameter,
$\dd_{\rm exp}$, is zero because of the Gutzwiller factor 
$g_\Delta =0$ at half filling, although $\dd$ and the variational parameter 
$\dd^V$ are nonzero.  
The self-consistent values which we obtain are (Fig.\ 10)
\beq
\mu^V  &&= 0, \nonumber\\
\dd^V  &&= \xx^V = 0.080 , \nonumber\\
\dd\ \ &&= \xx   = 0.163 , \nonumber\\
m^V    &&= 0.017 , \nonumber\\
m\ \ \ &&= 0.147.
\label{SCval}
\eeq
The actual expectation value with the projection, 
$m_{\rm exp}$, is (see Fig.\ 11)
\be
m_{\rm exp}= g_m m =0.373.
\ee
This is a reasonable magnitude and 
close to the Monte Carlo result for the 
Heisenberg model ($m_{\rm exp}=0.31 \pm 0.02$).\cite{Trivedi}  

The relations $\dd^V = \xx^V$ and $\dd = \xx$ are the manifestation 
of the SU(2) symmetry at half filling.\cite{Affleck,ZGRS}
Actually the self-consistency equations for $\xx$ and $\dd$ become 
the same at $\delta=0$.  
It is worth while noting here that, 
due to the SU(2) symmetry, the coexistent state of the AF and d-wave 
SC is equivalent to the $\pi$-flux state with AF long range order 
discussed by Hsu.\cite{Hsu}

The self-consistent value, $m = 0.147$ in Eq.\ (\ref{SCval}), 
is smaller than the result $m=1/2$ in the simple mean-field theory.  
Instead it is slightly larger than the value at which 
the Gutzwiller factor $g_s^Z$ has a maximum, as shown in Fig.\ 1.  
If $m$ is increased from the self-consistent value, $g_s^Z$ decreases.  
This indicates that $m$ is determined so as to minimize the exchange 
energy $-2NJ_{\rm eff}^Z m^2 = -2Ng_s^Z J m^2$ by optimizing 
the energy gain due to the long-range order and the energy loss 
due to the reduction of $g_s^Z$.  
This situation is, in some sense, similar to the 
quantum fluctuations of the Heisenberg spin system discussed 
in the spin-wave theory: 
The spin fluctuation, which reduces $m$ from the mean-field value $1/2$, 
leads to the enhancement of the Gutzwiller's renormalization factors 
to gain the energy.

(4) For less-than-half-filled case, 
the difference between the variational parameter $\dd^V$ and the 
expectation value $\dd_{\rm exp}$ is remarkable: 
$\dd^V$ is finite and moreover it increases as the doping rate 
$\delta$ decreases, 
while $\dd_{\rm exp}$ is proportional to $\delta$ near half filling.  
This is because the Gutzwiller factor, $g_\Delta$, 
is proportional to $\delta$ (see section VI).  
We interpret that $\dd^V$ is the BCS-type energy gap observed in 
scanning tunnel spectroscopy\cite{Renner,Oda,DeWilde,Matsuda,Nakano} 
or in break junctions\cite{Miyakawa},
because it is the parameter embedded in the wave function 
even at half filling.  
The excitation spectra will 
have a large energy gap corresponding to $\dd^V$, 
while the true long-range order, $\dd_{\rm exp}$, 
is reduced due to the projection or the strong correlation.  
The increase of $\dd^V$ in decreasing $\delta$ is consistent with the 
dependence observed experimentally.

(5) We discuss here the relation to the SO(5) theory.\cite{SO5}  
In our formulation, the combination $m^2 + \dd^2$ appears frequently.  
If the SO(5) symmetry is exact in the $t$-$J$ model, 
the free energy will have a systematic dependence such as
\be
F(\dd^2+m^2) \quad {\rm or} \quad F(\dd_{\rm exp}^2+m_{\rm exp}^2). 
\ee
For example, the numerator in $a$ has a combination
\be
16(m^2+\dd^2+\xx^2).
\ee
Also the factor giving the enhancement of $g_s^Z$  contains 
\be
4m^2+X_2 = 4m^2+2\dd^2+2\xx^2 .
\ee
Although these combinations remind us of the SO(5) symmetry,  
our Gutzwiller approximation does not show exact symmetry. 

Nevertheless a tendency similar to the SO(5) prediction can 
be seen from the Gutzwiller factors in Fig.\ 1.  
Comparing $g_s^{XY}$ and $g_s^Z$ for the cases with 
$\dd =0.02$ and $0.18$, we find that $g_s^{XY}$ and $g_s^Z$ are 
larger for smaller $\dd$.  
This is mainly from the exclusion effect of $a$, because even at $m=0$ 
this effect is observed.  
Therefore if we consider a situation where the d-wave SC order 
parameter is suppressed, then the Gutzwiller factor is enhanced 
causing the increase of AF moment.  
This gives the similar phenomena predicted in the SO(5) theory.\cite{Arovas}

(6) One advantage of the present theory is that it is easily applied 
to the inhomogeneous cases, such as the stripe state, vortex cores,
and magnetic states around nonmagnetic or magnetic impurities,
where the interplay between the AF and d-wave SC correlations plays 
an important role.  
Since our Gutzwiller approximation gives a reasonable estimate of 
the variational energies in the presence of AF correlations, 
a reliable analytic formulation can be given.  

The simplest way of applying the present scheme to inhomogeneous 
problems is to assume that $g_s^{XY}, g_s^Z$ and $g_t$ for each bond 
$\langle i,j \rangle$ are determined locally from the 
expectation values $\dd_{ij}, \xx_{ij}$ and $m_i, m_j$ for the bond.  
In this case, if the d-wave order parameter is reduced around 
vortex cores, impurities or stripes, then the Gutzwiller factors 
$g_s^{XY}$ and $g_s^Z$ are enhanced locally as expected from Fig.\ 1.  
This effect causes the local development of AF correlations, which 
can be observed experimentally.  
From these viewpoint, 
preliminary calculations for the vortex cores\cite{lasv,ISS99} 
and stripe states\cite{Yasuoka} have been published elsewhere.  

Of course the VMC simulations give more accurate evaluation of the 
variational energies, if they are used for the inhomogeneous systems 
like stripe states.  
However there are some difficulties in applying the VMC.  
Firstly we have to treat a fairly large unit cell to study the 
slowly varying order parameters especially near half filling.\cite{Kobayashi}
For example, 
the incommensurability in the stripe state is close to $(\pi, \pi)$ 
so that the period of the stripe pattern becomes fairly long.  
In this case the VMC simulations become difficult.  
Furthermore the choice of the functional form of the trial state 
in VMC is restricted, since only the small number of variational 
parameters can be used practically.  
On the contrary, our scheme based on the mean-field-type 
Gutzwiller approximation can be used in a fairly large system sizes.  
Moreover the order parameters on all the bonds, $\dd_{ij}, \xx_{ij}$ and 
$m_i$ can be optimized in the similar sense to unrestricted Hartree-Fock 
theory.  
Therefore we can search for the microscopically optimized variational 
states in our scheme.\cite{lasv,ISS99,Yasuoka}

The authors wish to thank T.\ M.\ Rice, H.\ Fukuyama, 
Y.\ Tanaka, H.\ Tsuchiura and M.\ Sigrist for useful discussions.  
They also thank C.-M.\ Ho for critical reading of the manuscript.  
This work is supported in part by a Grant-in-Aid of 
of the Ministry of Education, Science, Sports and Culture.

\appendix
\section{Derivation of the expectation value of an operator}
The evaluation of Eq.\ (\ref{Oexp}) is 
slightly complicated.  
The constraints for $\{ N_i'\}$ are
\beq
\sum_{i=1}^K n_{\uparrow i} N_i' &&= \frac{N_e}{2}-n_{\uparrow i_0}, 
\sum_{i=1}^K n_{\downarrow i} N_i' = \frac{N_e}{2}-n_{\downarrow i_0}, 
\nonumber\\
\sum_{i=1}^K \nhi N_i' &&= N- N_e -n_{{\rm h}i_0}, 
\eeq
instead of (\ref{constr}) because 
$N_i'$ represents the number of cells of the $i$-th state 
except for the central cell.  

In the same way as in the denominator, the largest term is given by 
\beq
\overline{N_i'}=\frac{N}{N_c}\frac{\omega_i}{W'} (p')^\nhi,
\eeq
with slightly different values of $W'$ and $p'$ 
because of the difference of the constraints.  
From the constraints (A1), 
we can see that $\overline{N_i'}$ satisfies the relations 
\be
\sum_{i=1}^K \overline{N_i'} = \frac{N}{N_c}-1, \quad
\sum_{i=1}^K \nhi \overline{N_i'} =N- N_e-n_{{\rm h}i_0} .
\ee
Therefore if we define 
the difference $\Delta\overline{N_i}=\overline{N_i'}-\overline{N_i}$, 
we have important relations
\beq
\sum_{i=1}^K \Delta \overline{N_i} = -1,  \qquad 
\sum_{i=1}^K \nhi \Delta \overline{N_i} =-n_{{\rm h}i_0} .
\eeq

By use of these, the ratio between the numerator and 
the denominator in Eq.\ (\ref{numegen}) is calculated as follows:
\beq
\widehat{\cal O} &&=
\frac{\langle \psi_0 | P_G \widehat{\cal O} P_G  | \psi_0\rangle }
{\langle \psi_0 | P_G P_G  | \psi_0\rangle } \nonumber\\
&&= \sum_{i_0} \frac{1}{\bigl(\frac{N}{N_c}\bigr)} 
\prod_{i=1}^K \frac{\overline{N_i}!}{\overline{N_i'}!}
\prod_{i=1}^K \omega_i^{\overline{N_i'}-\overline{N_i}} 
\langle \widehat{\cal O} \rangle_{i_0} \nonumber\\
&&= \sum_{i_0} \frac{N_c}{N}
\prod_{i=1}^K \biggl( \frac{\overline{N_i}}{\omega_i} 
\biggr)^{-\Delta \overline{N_i}} 
\langle \widehat{\cal O} \rangle_{i_0} \nonumber\\
&&= \sum_{i_0} \frac{N_c}{N}
\prod_{i=1}^K \biggl( \frac{N}{N_c}\frac{p^\nhi}{W}
\biggr)^{-\Delta \overline{N_i}} 
\langle \widehat{\cal O} \rangle_{i_0} \nonumber\\
&&= \sum_{i_0} \frac{N_c}{N}
\biggl( \frac{N}{N_c}\frac{1}{W} \biggr)^{-\sum \Delta \overline{N_i}} 
\times p^{-\sum \nhi \Delta \overline{N_i}} 
\langle \widehat{\cal O} \rangle_{i_0} \nonumber\\
&&= \sum_{i_0}\frac{p^\nhiz}{W} \langle \widehat{\cal O} \rangle_{i_0}.
\eeq
This gives Eq.\ (\ref{numegen}).

\section{Reproduction of the original Gutzwiller approximation}

In this appendix we show that the 
original Gutzwiller approximation described in II is reproduced 
if we assume only the site-diagonal expectation values 
in the generalized formulation in section III.  

First we calculate the weight $\omega_i$ of the $i$-th state which 
has $n_{\uparrow i}$ up-spin electrons, $n_{\downarrow i}$ down-spin 
electrons and $\nhi$ holes.  If we take only the site-diagonal expectation 
values, we have
\be
\omega_i= [r(1-w)]^{n_{\rm right}} [w(1-r)]^{n_{\rm wrong}}
[(1-r)(1-w)]^\nhi,
\ee
where $n_{\rm right}$ ($n_{\rm wrong}$) means the number of sites  
where the right (wrong) spices of spin direction is located 
depending on the sublattice 1 and 2.  
Then the total weight in the subgroup with 
$j$ holes (Eq.\ (\ref{wj})) can be calculated exactly as
\beq
W_j &&= \sum_{i\ {\rm with}\ j \ {\rm holes}} \omega_i \nonumber\\
&&= \ _{N_c} C_j [r(1-w)+w(1-r)]^{N_c-j} \left[ (1-r)(1-w)\right]^j \nonumber\\
&&= \ _{N_c} C_j (n-2rw)^{N_c-j} (1-r)^j (1-w)^j ,
\eeq
where $\ _{N_c} C_j$ is the number of choices of the positions 
of $j$ holes.  

Using these $W_j$, the constraints (\ref{constr3}) have simple forms as
\beq
&&\sum_{j=0}^{N_c} \frac{W_j}{W} p^j =\frac{1}{W} 
\{ n-2rw +(1-r)(1-w)p\}^{N_c}= 1, \nonumber\\
&&\sum_{j=0}^{N_c} j\frac{W_j}{W} p^j =\frac{N_c}{W} (1-r)(1-w)p \nonumber\\
&&\qquad \ \times \{ n-2rw +(1-r)(1-w)p\}^{N_c-1}= \MYdelta N_c.
\eeq
These can be solved easily to give
\beq
p=\frac{\delta (n-2rw)}{n(1-r)(1-w)}, \nonumber\\
W=\biggl( \frac{n-2rw}{n} \biggr)^{N_c}.
\eeq
Finally the expectation value for $S_\ell^+ S_m^-$, for example,
becomes
\beq
&&\frac{\langle \psi_0 | P_G S_\ell^+ S_m^- P_G  | \psi_0\rangle }
{\langle \psi_0 | P_G P_G  | \psi_0\rangle } \nonumber\\
&&\ = \sum_{i_0}\frac{p^\nhiz}{W} 
\langle S_\ell^+ S_m^- \rangle_{i_0} \nonumber\\
&&\ = \sum_{j=0}^{N_c-2} \frac{p^j}{W} \ _{N_c-2}C_j 
(n-2rw)^{N_c-2-j} (1-r)^j (1-w)^j \langle S_\ell^+ S_m^- \rangle_0 \nonumber\\
&&\ = \frac{1}{W} \{ n-2rw +(1-r)(1-w)p\}^{N_c-2} 
\langle S_\ell^+ S_m^- \rangle_0 \nonumber\\
&&\ = \frac{n^2}{(n-2rw)^2} \langle S_\ell^+ S_m^- \rangle_0,
\eeq
which is exactly the same as the results in 
the original Gutzwiller approximation in section II.B, Eq.\ (\ref{gs0}).

\section{Evaluation of $g_s^Z$ for the less-than-half-filled case}

In the similar way to $g_s^{XY}$, we evaluate $g_s^Z$.  
In the zero-hole sector, we have
\beq
\sum_{i_0\ {\rm with}\ 0\ {\rm holes}} \langle S_\ell^z S_m^z \rangle_{i_0}
&&= \langle S_\ell^z S_m^z \rangle_0 (n-2rw)^{N_c-2} a^{\~{N_b}} \nonumber\\
-&&2N_2 m^2 X_2 (n-2rw)^{N_c-3} a^{\~{N_b}-3} \nonumber\\
-&&N_2^2 m^2 X_2^2 (n-2rw)^{N_c-4} a^{\~{N_b}-6} . \nonumber \\
&&\ 
\eeq
For the one-hole sector, there are five contributions from the 
diagrams in Fig.\ 9.  The first three diagrams are simple extensions 
of the zero-hole sector.  However Figs.\ 9(d) and 9(e) are new-type 
contributions due to the presence of a hole.  
For the diagrams in Figs.\ 9(d) and 9(e) we calculate 
\beq
\pmatrix{\ && \circ \cr \bullet && \bullet} 
&&=\langle 
S_\ell^z S_m^z (1-\widehat n_{m'\uparrow})(1-\widehat n_{m'\downarrow})
\rangle_{c} \nonumber\\
&&=\frac{1}{2}m^2X_2, \nonumber\\
\pmatrix{\ && \circ \cr \bullet && \bullet \cr \bullet && \ } 
&&=\langle 
\bigl\{
\widehat n_{\ell'\uparrow}(1-\widehat n_{\ell'\downarrow})+
\widehat n_{\ell'\downarrow}(1-\widehat n_{\ell'\uparrow})\bigr\}
S_\ell^z S_m^z \nonumber\\
&&\quad \times (1-\widehat n_{m'\uparrow})(1-\widehat n_{m'\downarrow})
\rangle_{c} \nonumber\\
&&=\frac{1}{2}m^2X_2^2.
\eeq
Using these expectation values, we obtain
\beq
&&\sum_{i_0\ {\rm with}\ 1\ {\rm holes}} \langle S_\ell^z S_m^z \rangle_{i_0}
\nonumber\\
&&\qquad = _{N_c-2}C_1 \langle S_\ell^z S_m^z \rangle_0 z (n-2rw)^{N_c-3} 
a^{\~{N_{1b}}} \nonumber\\
&&\qquad\ -_{N_c-3}C_1 2N_2 m^2 X_2 z (n-2rw)^{N_c-4} a^{\~{N_{1b}}-3}
\nonumber\\
&&\qquad\ -_{N_c-4}C_1 N_2^2 m^2 X_2^2 z (n-2rw)^{N_c-5} 
a^{\~{N_{1b}}-6} \nonumber\\
&&\qquad\ + N_2 m^2 X_2 (n-2rw)^{N_c-3} a^{\~{N_{b}}-3} \nonumber\\
&&\qquad\ + N_2^2 m^2 X_2^2 (n-2rw)^{N_c-4} a^{\~{N_{b}}-6}.
\eeq

In the two hole sector, we need to calculate a diagram in Fig.\ 9(f) which 
is $-\frac{1}{4} m^2X_2^2$.  Taking account of these diagrams and 
counting the higher order terms with respect to $X$, we approximate as
\beq
&&\sum_{i_0\ {\rm with}\ 2\ {\rm holes}} \langle S_\ell^z S_m^z \rangle_{i_0}
\nonumber\\
&&\qquad = _{N_c-2}C_2 \langle S_\ell^z S_m^z \rangle_0 z^2 (n-2rw)^{N_c-4} 
a^{\~{N_{2b}}} \nonumber\\
&&\qquad\ - _{N_c-3}C_2 2N_2 m^2 X_2 z^2 (n-2rw)^{N_c-5} a^{\~{N_{2b}}-3} 
\nonumber\\
&&\qquad\ - _{N_c-4}C_2 N_2^2 m^2 X_2^2 z^2 (n-2rw)^{N_c-6} a^{\~{N_{2b}}-6}
\nonumber\\
&&\qquad\ + _{N_c-3}C_1 N_2 m^2 X_2 z (n-2rw)^{N_c-4} a^{\~{N_{1b}}-3} 
\nonumber\\
&&\qquad\ + _{N_c-4}C_1 N_2^2 m^2 X_2^2 z (n-2rw)^{N_c-5} a^{\~{N_{1b}}-6} 
\nonumber\\
&&\qquad\ -\frac{N_2^2}{4} m^2 X_2^2 (n-2rw)^{N_c-4} a^{\~{N_{b}}-6}.
\eeq

Finally generalization to higher order contributions leads to
\beq
&&\langle S_\ell^z S_m^z \rangle = \frac{1}{W} \sum_{j=0}^{N_c} p^j 
\sum_{i_0\ {\rm with}\ j\ {\rm holes}} 
\langle S_\ell^z S_m^z \rangle_{i_0} \nonumber\\
&&= \frac{1}{W} \sum_{j=0}^{N_c-2} p^j 
\ _{N_c-2}C_j \langle S_\ell^z S_m^z \rangle_0 z^j(n-2rw)^{N_c-2-j} 
a^{\~{N_{j,b}}} \nonumber\\
&&\ -\frac{1}{W} \sum_{j=0}^{N_c-3} p^j 
\ _{N_c-3}C_j 2N_2 m^2 X_2 z^j (n-2rw)^{N_c-3-j} a^{\~{N_{j,b}}-3} \nonumber\\
&&\ -\frac{1}{W} \sum_{j=0}^{N_c-4} p^j 
\ _{N_c-4}C_j N_2^2 m^2 X_2^2 z^j (n-2rw)^{N_c-4-j} a^{\~{N_{j,b}}-6}\nonumber\\
&&\ +\frac{1}{W} \sum_{j=0}^{N_c-3} p^{j+1}
\ _{N_c-3}C_j N_2 m^2 X_2 z^j (n-2rw)^{N_c-3-j} a^{\~{N_{j,b}}-3} \nonumber\\
&&\ +\frac{1}{W} \sum_{j=0}^{N_c-4} p^{j+1} 
\ _{N_c-4}C_j N_2^2 m^2 X_2^2 z^j (n-2rw)^{N_c-4-j} a^{\~{N_{j,b}}-6}
\nonumber\\
&&\ -\frac{1}{W} \sum_{j=0}^{N_c-4} p^{j+2} 
\ _{N_c-4}C_j \frac{N_2^2}{4} m^2 X_2^2 z^j (n-2rw)^{N_c-4-j} a^{\~{N_{j,b}}-6}
\nonumber\\
&&= \biggl( \frac{n}{n-2rw}\biggr)^2 \langle S_\ell^z S_m^z \rangle_0 
a^{-(N_b-\~{N_{b}})} \nonumber\\
&&\ - \biggl( \frac{n}{n-2rw}\biggr)^3  
 2N_2 m^2 X_2 \bigl( 1-\frac{p}{2}\bigr) a^{-(N_b-\2~{N_{b}})} \nonumber\\
&&\ - \biggl( \frac{n}{n-2rw}\biggr)^4  
 N_2^2 m^2 X_2^2 \bigl( 1-\frac{p}{2}\bigr)^2 a^{-(N_b-\3~{N_{b}})} \nonumber\\
&&= \biggl( \frac{n}{n-2rw}\biggr)^2 a^{-(N_b-\~{N_{b}})}  \nonumber\\
&&\qquad \times \biggl[ -\frac{X_2}{4} - m^2\bigl\{ 1+\frac{N_2 X_2 n}{n-2rw}
\bigl( 1-\frac{p}{2}\bigr) a^{-3} \bigr\}^2 \biggr].
\eeq
Thus the Gutzwiller factor is
\be
g_s^Z = g_s^{XY} \frac{1}{4m^2+X_2} 
\biggl[ X_2 + 4m^2\bigl\{ 1+\frac{N_2 X_2 n}{n-2rw} \bigl(1-\frac{p}{2}\bigr)
a^{-3} \bigr)^2 \biggr].
\ee

% now the references. delete or change fake bibitem. delete next three
%   lines and directly read in your .bbl file if you use bibtex.
\def\journal#1#2#3#4{#1 {\bf #2} (#4) #3}
\def\PR{Phys.\ Rev.}
\def\PRB{Phys.\ Rev.\ B}
\def\PRL{Phys.\ Rev.\ Lett.}
\def\PL{Phys.\ Lett.}
\def\JPSJ{J.\ Phys.\ Soc.\ Jpn.}
\def\PTP{Prog.\ Theor.\ Phys.}
\def\SSC{Solid State Commun.}
\def\JPC{J.\ Phys.\ C}
\def\JETP{J.\ Exptl.\ Theoret.\ Phys.}
\def\SJETP{Sov.\ Phys.\ JETP}
\def\PLA{Phys.\ Lett.\ A}
\def\RMP{Rev.\ Mod.\ Phys.}
\def\JMP{J.\ Math.\ Phys.}
\def\IJMP{Int.\ J.\ Mod.\ Phys.}

%\end{thebibliography}

\end{document}